\documentclass[draft]{agujournal2019}
\usepackage{url} 
\usepackage{lineno}
\usepackage{soul}
\draftfalse

\journalname{Space Weather}

\begin{document}

%
%


\title{OSPREI: A Coupled Approach to Modeling CME-Driven Space Weather with Automatically-Generated, User-Friendly Outputs}

%
%




\authors{C. Kay\affil{1,2}, M. L. Mays\affil{1}, Y. M. Collado-Vega\affil{1}}

\affiliation{1}{Heliophysics Science Division, NASA Goddard Space Flight Center, Greenbelt, MD, USA}
\affiliation{2}{Dept. of Physics, The Catholic University of America, Washington DC, USA}






\correspondingauthor{Christina Kay}{christina.d.kay@nasa.gov}




\begin{keypoints}
\item OSPREI performs ensemble simulations of the Sun-to-Earth behavior of CMEs on time scales relevant for future forecasting 
\item A standardized set of automatically-generated visualizations can provide essential information for forecasters in an intuitive manner
\item Mimicking a forecasting approach, we apply OSPREI to the 2021 April 21 and 2021 May 09 CMEs to illustrate its capabilities
\end{keypoints}

%
%

%
%


\begin{abstract}
Coronal Mass Ejections (CMEs) drive space weather activity at  Earth and throughout the solar system. Current CME-related space weather predictions rely on information reconstructed from coronagraphs, sometimes from only a single viewpoint, to drive a simple interplanetary propagation model, which only gives the arrival time or limited additional information. We present the coupling of three established models into OSPREI (Open Solar Physics Rapid Ensemble Information), a new tool that describes Sun-to-Earth CME behavior, including the location, orientation, size, shape, speed, arrival time, and internal thermal and magnetic properties, on the timescale needed for forecasts. First, ForeCAT describes the trajectory that a CME takes through the solar corona. Second, ANTEATR simulates the propagation, including expansion and deformation, of a CME in interplanetary space and determines the evolution of internal properties via conservation laws. Finally, FIDO produces in situ profiles for a CME’s interaction with a synthetic spacecraft. OSPREI includes ensemble modeling by varying each input parameter to probe any uncertainty in their values, yielding probabilities for all outputs. Standardized  visualizations are automatically generated, providing easily-accessible, essential information for space weather forecasting. We show OSPREI results for a CMEs observed in the corona on 2021 April 22 and 2021 May 09. We approach these CME as a forecasting proof-of-concept, using information analogous to what would be available in real time rather than fine-tuning input parameters to achieve a best fit for a detailed scientific study. The OSPREI ``prediction'' shows good agreement with the arrival time and in situ properties. 
\end{abstract}

\section*{Plain Language Summary}
Coronal mass ejections (CMEs) are large plasma structures that erupt from the Sun. These eruptions occur frequently but, depending on their exact properties (such as speed and magnetic field strength and orientation), can have negative effects on human technologies both in space or on the surface of the Earth. It is important to know where a CME is going, when it will arrive, and its properties upon impact. Currently, most CME forecasts are limited to the time of arrival, and potentially the speed at that time, but have little to no additional information. We present OSPREI, which couples existing models to simulate the full Sun-to-Earth evolution of a CME. OSPREI simulates the change in the path of a CME near the Sun, a range of effects during it's transit to Earth, and what it would look like as it passes over a spacecraft. OSPREI runs very quickly so we can see how results change as the inputs change. To facilitate forecasts, we have developed a set of standardized figures that present OSPREI results (including CME location, time of arrival, and internal properties) in an easily accessible format displaying the most likely values and the range of possibilities.

%
%

%


%
%
%
%

\section{Introduction}
Mankind has noticed the effects of space weather on human technologies for centuries. These changes in the near-Earth plasma, radiation, and charged particle environment has plagued technology from magnetic compasses to telegraphs, but scientific interest rocketed at the dawn of the space age when satellites began both sampling the near-Earth environment and being disrupted by it. Forecasting space weather has gained increased importance in the public eye with recent emphasis from NASA's Moon to Mars initiative and the US government's National Space Weather Strategy and Action Plan in 2019 and the Promoting Research and Observations of Space Weather to Improve the Forecasting of Tomorrow (PROSWIFT) Act in 2020. Coronal mass ejections (CMEs), huge explosions of plasma and magnetic field that commonly erupt from the Sun, are key drivers of space weather, both near the Earth and throughout the rest of the solar system, understanding and forecasting their behavior should be a part of any space weather prediction. Currently, most predictions focus simply on the arrival time of a CME. The severity of CME-driven space weather effects depends on the actual properties of the CME, such as its magnetic field strength and orientation. Predicting the magnetic field remains an open challenge, most arrival time predictions only include the hydrodynamic properties, such as density and velocity, and an estimate of the maximum Kp value, which is an index describing the severity of a geomagnetic storm.

The first step to determining the potential impact of a CME is to determine where it exists in three-dimensional space. Many eruptions have a visible source on the solar disk, but some are back-sided or even ``stealth'' CMEs with no obvious signatures on disk or in the low corona \cite<e.g.>{Rob09,Oka19}. As the CME erupts and begins propagating through the corona it may deflect or rotate, changing its position or orientation from those initial values \cite<e.g.>{Byr10,Gui11}. Typically, a CME's position is reconstructed by visually fitting a geometric shape to one or more coronagraph images \cite<e.g.>{Xie04,The06}. This yields information about a CME's position and size, and potentially its orientation. While relatively easy to routinely apply to observations, there can be significant uncertainties in the reconstructed values \cite<e.g.>{Mie10}. The technique, however, allows for a quick estimate on whether a CME's path will cause an impact at the Earth or at another location of interest \cite<e.g>{Rod11}.

At the most basic level, if the trajectory is known, then a speed can be calculated from a height-time profile, and an arrival time estimated using simply the distance and speed, but this method leads to large errors as it neglects the change in CME speed due to drag. Any worthwhile forecasting model must include some form of drag interaction between the CME and background solar wind. Analytical drag models can be as simple as a one-dimensional \cite<e.g>{Vrs07DBM}, or incorporate the geometry and size of the CME \cite<e.g.>{Zic15}. More sophisticated, three-dimensional models embed a CME pulse in a solar wind background and numerically solve a set of fluid or magnetohydrodynamic (MHD) differential equations over the entire simulation domain, yielding not only the arrival time but the evolution of the CME itself \cite<e.g.>{Ods99,Ril12,Shi16,Pom18}. While more computationally expensive than the analytic models, these fluid models can be used for forecasts, but real-time results tend to be restricted to individual runs or restricted to a small number of ensemble runs (see discussion of real-time ensembles in \citeA{May15AT}), limiting any information on the probabilities of specific outcomes. \citeA{Wol18} found a mean absolute error of 10 hours for the real-time forecasts of 273 CMEs from the Community Coordinated Modeling Center using the WSA-ENLIL+Cone model. This value tends to be representative of the error for most current arrival time models \cite<e.g.>{Ril18}.

Addressing the severity of the impact requires knowing the CME properties. Most arrival time models also produce the speed of a CME as it is intrinsically related to any physics-based calculation of transit time. The fluid models include predictions of the density of the CME, but typically the CME is treated as a pressure pulse and does not have an internal magnetic structure. Very sophisticated, fully-magnetohydrodynamic simulations are possible \cite<e.g>{Tor18, Ver19euh}, which can provide great details for individual scientific studies \cite{Asv21,Sco21a,Sco21b}. Some MHD models can be run as ensembles, even those including a magnetic structure for the CME (rather than simply a pressure pulse) but when the models include the coronal evolution of a CME they tend to become difficult to run in real time.  Several analytic models exist that include some evolution of internal CME properties at either coronal or interplanetary distances \cite{Kum96,Dur17,Mis18}, but these have not yet been incorporated into forecasts.

Alternatively, the magnetic structure of a CME can be forward modeled by combining the observed position of a CME with a flux rope model, allowing for the generation of synthetic in situ profiles \cite<e.g.>{Kun10,Sav15}. This method allows for rapid results but does rely on many simplifying assumptions and can only produce the general trends and not any smaller-scale variation. Despite the relative simplicity of this method it is not yet routinely used in forecasts.

\citeA{Kay1745} begun working toward a unified Sun-to-Earth CME forecasting model, which led to the Open Solar Physics Rapid Ensemble Information (OSPREI) model presented in this paper. This work combined Forecasting a CME's Altered Trajectory \cite<ForeCAT,>{Kay13,Kay15}, a model for the coronal deflection and rotation of CMEs, with the ForeCAT In situ Data Observer \cite<FIDO>{Kay17FIDO}. ForeCAT provided the position and orientation of a CME, which at that point was assumed to arrive at the synthetic spacecraft of FIDO at the observed time of arrival. No interplanetary evolution was included at this time. \citeA{Kay1745} used the combination of ForeCAT and FIDO to reproduce 45 CMEs observed in situ near Earth and remotely by both STEREO spacecraft. \citeA{Kay18} introduced an ensemble approach so that the sensitivity to inputs could be explored by randomly varying their values within the range of their estimated uncertainties. At the same time, ANother Type of Ensemble Arrival Time Results (ANTEATR) was introduced, adding a simple drag-based arrival time model to fill the interplanetary gap. In \citeA{Kay18}, the three models were not directly coupled but ran individually with results from one manually passed to the next model in the chain. Subsequently, the ability to include a CME-driven sheath has been added as FIDO-Sheath Induced by Transient \cite<FIDO-SIT>{Kay20SIT} and internal magnetic and thermal forces added to ANTEATR as ANTEATR-Physics Driven Approach to Realistic Axis Deformation and Expansion \cite{KayAP1,KayAP2}, which now simulates the expansion and deformation of a CME instead of assuming constant values as before. \citeA{KayAP1} presents the theoretical background of the new model and analyzes the relative importance of drag, magnetic, and thermal forces and \citeA{KayAP2} determines the sensitivity of the results, such as transit time, duration, final propagation and expansion velocities, and internal magnetic field and temperature, to various input parameters. 

Now, for the first time, we present a fully-coupled suite using the most recent versions of ForeCAT, ANTEATR, and FIDO and have developed a set of visualizations designed to facilitate space weather forecasting. We apply OSPREI to two observed CMEs as if they were being forecast, and describe the resulting ``predictions.'' We emphasize that we are mimicking a forecasting approach and only using information that would be available almost immediately after eruption, days before the CME arrives at Earth. This work highlights the range of information that could potentially be made available to a forecaster in the near future.

\section{OSPREI Suite}\label{model}
\begin{figure}
 \noindent\includegraphics[width=\textwidth]{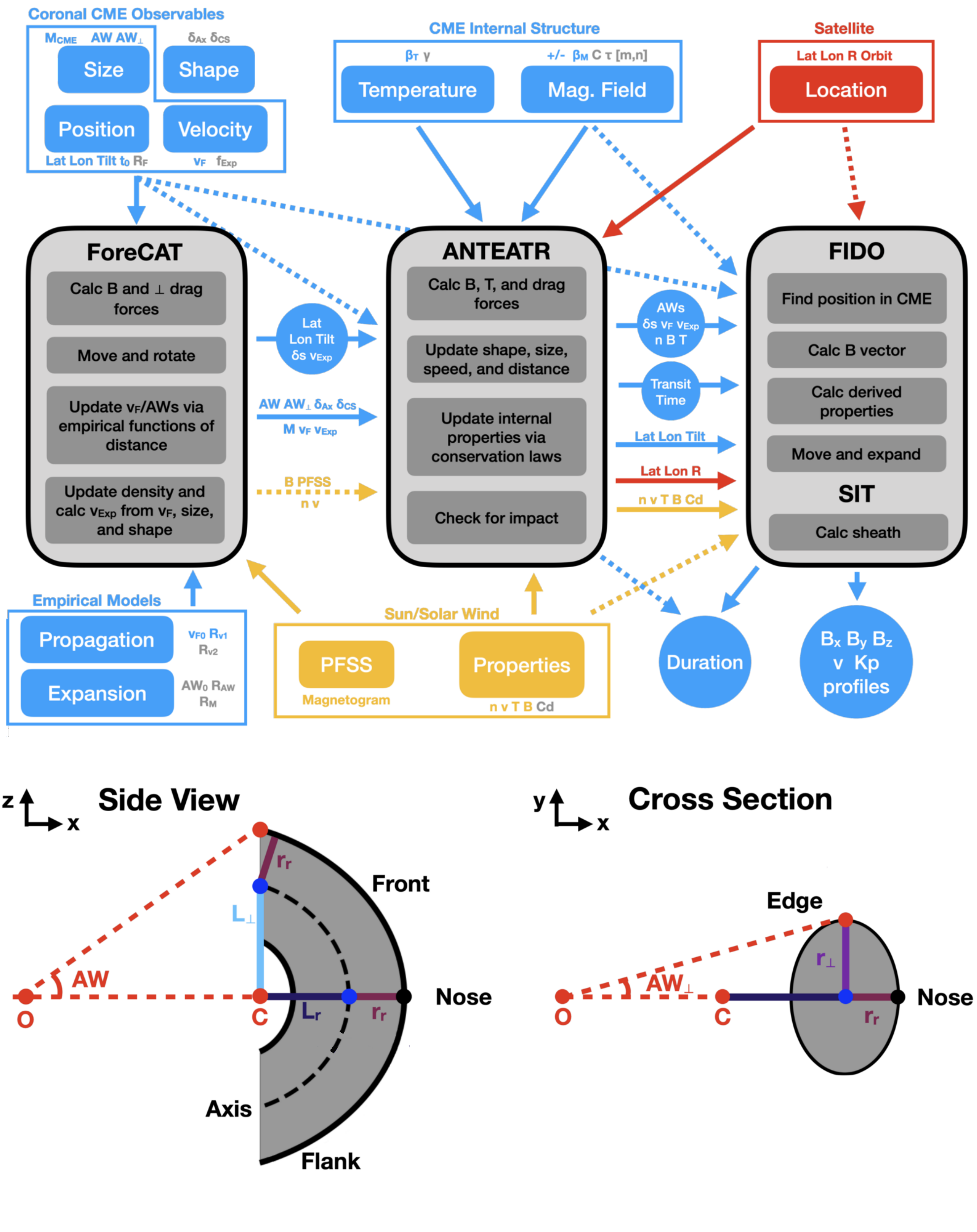}
\caption{Schematic showing the three components of OSPREI and the connections between their inputs and outputs (top). Cartoon showing the CME shape from the side and through the cross section (bottom).}
\label{OSP}
\end{figure}

OSPREI combines ForeCAT, ANTEATR, and FIDO into a fully-coupled tool to facilitate potential forecasts. Each of these models has been thoroughly described in their individual papers so here we only provide a brief summary and focus on the flow of information between them. Figure \ref{OSP} contains a schematic showing the three components of OSPREI, the connections between them, and their individual inputs and outputs. The three gray regions represent the three individual models and give a basic algorithm for each one. The smaller, rounded rectangles represent model inputs and are colored according to CME (blue), background solar wind (yellow), and satellite (red) inputs. We give general descriptions of the inputs within the rectangles (such as ``Position'') but include the specific variable names nearby (such as ``Lat''). We expect that variables in gray would typically be left at default values during a forecast but the colored ones would need to be measured or estimated from observations.  The circles correspond to outputs from each model. Solid arrows show the flow of information if one runs all three components of OSPREI. The dashed arrows show alternative paths if one or more of the individual components is not included. For example, if ForeCAT is not run because one wanted to use inputs from a coronal reconstruction instead, then the CME position would become an input for ANTEATR. When all three components are used, outputs from one model become inputs for the next and the paths indicated by dashed arrows are not used.

The bottom of Fig. \ref{OSP} includes a cartoon showing the CME shape used by all three components, with the left showing a side view and the right a slice through a cross section. While the early version of ForeCAT use a toroidal axis defined by an ellipse with semimajor and semiminor lengths $L_{\perp}$ and $L_r$, \citeA{KayAP1} find that this shape produces unnaturally large curvature and therefore magnetic tension forces near the flanks. A parabolic axis running through the points defined by the central point C and $L_r$ and $L_{\perp}$ is more well-behaved near the flanks but has excessively high curvature at the nose. By averaging the elliptical and parabolic axis, \citeA{KayAP1} created a new toroidal axis with well-behaved curvature throughout. \citeA{KayAP1} also introduced an elliptical cross section to the CME torus, defined by the lengths $r_{\perp}$ and $r_r$, which are assumed to be uniform along the entire toroidal axis. The full CME surface can be specified by the radial distance of the nose and the four lengths $L_r$, $L_{\perp}$, $r_r$, and $r_{\perp}$, or the full and cross-sectional angular widths, $AW$ and $AW_{\perp}$, and the aspect ratios $\delta_{Ax}=L_r / L_{\perp}$ and $\delta_{CS}=r_r / r_{\perp}$. In practice, the angular widths and aspect ratios tend to be more intuitive and easier to measure or estimate than the individual lengths. 

The first model in the OSPREI chain is ForeCAT, which simulates the deflection and rotation of a CME through the corona. The initial CME size and shape are set via the initial distance, the two angular widths and aspect ratios, and the CME latitude, longitude, and tilt. The deflection and rotation are determined from the forces from the background solar magnetic field, so the CME mass is need to convert these forces to accelerations. The net force from magnetic tension and magnetic pressure gradients creates a change in the CME's latitude and/or longitude and the net torque from the deflection forces produces a rotation about the CME nose, changing its tilt.

While the deflection and rotation are simulated from physics, the CME's radial propagation, expansion, and mass evolution are specified via empirical models. ForeCAT is intended to simulate from the earliest observed upward motion of a CME so it begins with a slow rise phase at a constant, slow velocity, followed by rapid acceleration, and finally constant propagation (until it reaches ANTEATR where drag is included). The empirical coronal propagation model is set by the slow rise velocity, and the two radii at which the phase transitions occur ($R_{v1}$ is slow rise to rapid acceleration, $R_{v2}$ is rapid acceleration to constant propagation). The empirical expansion model replicates a rapid overexpansion as the CME erupts using the form $AW = AW_0 + \Delta AW(1-\exp(-(R_F-1) / R_{AW})$, where $R_F$ is the distance of the CME front, $R_{AW}$ is a length scale controlling the rate of expansion, $AW_0$ is the angular width for a $R_F$ of 1 $R_S$ (approximately but not exactly the initial angular width), and $\Delta AW$ is the difference between the maximum coronal $AW$ and $AW_0$. We also include an empirical increase in the CME mass to mimic the pileup of coronal material. The CME begins at half of the input value and linearly increases until it reaches a user-specified distance, $R_M$. Typically we leave these empirical model parameters at default values unless we have a good reason to modify them based upon observations.

ForeCAT relies on a Potential Field Source Surface model \cite<PFSS, e.g.>{Alt69} of the background coronal magnetic field with a source surface height of 2.5 $R_s$, which we initialize using a synchronic magnetogram corresponding to the date of the eruption. We use the ``Daily Update Radial Synoptic Map'' from the Solar Dynamic Observatory's Helioseismic and Magnetic Imager (SDO HMI), available as ``hmi.Mrdailysynframe\_720s'' at the Joint Science Operations Center (JSOC). We automatically identify the location of the heliospheric current sheet (HCS) from the PFSS results, which drives a background solar wind density and velocity model based on \citeA{Guh06}. 

The main outputs of ForeCAT are the latitude, longitude, and tilt of the CME in the outer corona (user-set distance but typically within 10-20 $R_S$ heliocentric distance). Previous work has shown that CMEs tend to deflect away from coronal holes toward the HCS and rotate toward having their toroidal axis aligned with the HCS \cite<e.g>{Kay15L,Kay17AR}. ForeCAT also passes along the AWs, aspect ratios, and maximum coronal velocity, but these should not be thought of as outputs as they are not modified from their designated input values.

Next, ANTEATR, takes the output position and orientation from ForeCAT, as well as the size, shape, mass, and propagation velocity that have been forward along from the inputs. A new input $f$ sets the initial velocity decomposition (IVD), which sets how the front velocity, $v_F$ is broken down into a propagation velocity (the radial movement of ``C'' in Fig. \ref{OSP}) and expansion speeds (which relate to the change in the $L$ and $r$ lengths). We consider two extremes for the IVD, either fully self-similar or fully convective. For the self-similar case, both the angular widths and the aspect ratios remain constant as the CME propagates radial. For the convective case, the full front of the CME moves out in the local radial direction at $v_F$, which maintains constant angular widths but the aspect ratios decrease as the CME ``pancakes,'' becoming wider in the perpendicular direction than the radial direction. This value $f$ only affects the conversion of $v_F$ into the individual expansion velocities at the first time step of ANTEATR, at all subsequent times the velocities evolve according to the model forces.

OSPREI uses the newest version, ANTEATR-PARADE, which includes the external drag from the background solar wind, as well as the forces from the internal CME magnetic field and temperature. These inputs were not needed by ForeCAT but must be specified for ANTEATR. We use the elliptic-cylindrical (EC) flux rope model of \cite{Nie18}, which depends on parameters $\tau$, setting the internal twist distribution, and $C$, which sets the ratio of toriodal to poloidal magnetic field. We use the simplest version of the EC model with the polynomial exponents [m,n] set to [0,1]. While these parameters are specific to the EC flux rope model, and not likely to be adjusted in a forecasting scenario, any magnetic field model will require the magnetic field strength. We set the toroidal magnetic field at the center of the toroidal axis as $\beta_M$ times the background magnetic field at that location at the start of the simulation. Similarly, we set the internal temperature, which is approximated as uniform within the flux rope, as $\beta_T$ times the background solar wind temperature. Our temperature model also requires the adiabatic index, $\gamma$, which can be set anywhere between isothermal (1) and adiabatic (5/3).

At this point additional inputs describing the background solar wind must also be incorporated. We assume a very simple background solar wind model with constant velocity, density falling as $R^2$, a Parker spiral magnetic field, and temperature proportional to $R^{-0.58}$ \cite{Hel13}. These simple profiles can be scaled by either the outer coronal values from ForeCAT or their values at 1 AU, either from in situ observations or taken from a solar wind forecast. If given as inputs, OSPREI will prioritize any 1 AU inputs over the ForeCAT solar wind values. We find that this tends to be a more realistic background solar wind and more accurate arrival time results due to some of the shortcomings of the PFSS model. 

ANTEATR also requires the location (latitude, longitude, and distance) and orbital speed of the Earth or satellite of interest. Alternatively, it can be given a text file containing the location of a satellite as a function of time, as was done recently for an OSPREI simulation of a CME that impacted Parker Solar Probe \cite{Pal21PSP}.

ANTEATR calculates the drag between the CME and the background solar wind, which affects the full CME, the hoop force and toroidal magnetic tension, which affect the toroidal axis, and the poloidal magnetic tension and toroidal magnetic pressure and thermal pressure gradients (relative to the background solar wind), which affect the cross section. These forces modify the propagation and expansion velocities, giving a CME's distance as a function of time, but also the change in the angular widths and aspect ratios due to the changes in the $L$ and $r$ lengths. As the CME evolves the internal properties are updated via conservation laws so that the outputs include the internal magnetic field strength, temperature, and density. We stop the simulation once the CME impacts the satellite but approximate the duration based on the CME geometry and speed, and determine an estimated Kp.

Typically, we find that the angular widths remain constant or slowly decrease and that the CME pancakes, or flattens in the radial direction relative to the perpendicular direction. Whatever the values may be, we pass the arrival time, evolved velocities, angular widths, aspect ratios, and internal properties from ANTEATR to FIDO, as well as the position and orientation from ForeCAT (which do not change within ANTEATR). The only new input needed by FIDO is the handedness of the CME flux rope, which is not technically required for any of the ANTEATR force calculations.

FIDO determines the relative position of the spacecraft to the CME and calculates the magnetic field vector and velocity for that specific location. It also estimates the Kp index from the speed and magnetic field as done in \citeA{Kay20SIT}. Within FIDO, the CME continues to expand as it passes over the spacecraft according to the speeds provided by ANTEATR. If desired, FIDO also simulates the shock and sheath ahead of a CME using the Rankine-Hugoniot shock conditions with the CME and solar wind properties \cite{Kay20SIT}. This model calculates the compression of the magnetic field strength and density, and the velocity at the CME-driven shock and connects these values to the flux rope values to generate profiles within the sheath. This is a simple analytic shock model for the average sheath properties and cannot produce the stochastic fluctuations that are commonly observed. 

Within OSPREI these components are fully coupled so that a single call to the model runs the full Sun-to-Earth simulation. OSPREI can also be set to run an ensemble where the user specifies the number of individual runs, the input parameters to randomly vary, and their ranges. To generate the ensemble, a new input value is generated from a normal distribution centered at the seed value and with the standard deviation set to one-third of the specified range (making 99.7\% of randomly chosen values within the range). All specified inputs are varied simultaneously so that if five inputs are to be varied then all other parameters will be held constant but each five of those will differ in all ensemble members. Standardized data files are created, which can then be immediately processed into user-friendly figures using a separate routine. We also note that combinations of the various components of OSPREI can be run instead of the full set. For example, one can specify the coronal position and orientation from a coronal reconstruction and use those values and uncertainties to drive ANTEATR and FIDO instead of the ForeCAT simulation. The full OSPREI source code is publicly available to the community via Zenodo and GitHub. OSPREI has also been delivered to the Community Coordinated Modeling Center (CCMC). CCMC provides free access to heliophysics simulations to the international science community and supports the transition of models from research to operations.  Soon users will be able to request OSPREI simulations via CCMC's Runs-on-Request and Instant Run services.  OSPREI has begun CCMC model onboarding, which is a collaborative process that involves installation, testing, defining metadata, creating a run submission interface, and designing a run output page (\url{https://ccmc.gsfc.nasa.gov/models/model\_on\_board.php}).

Additionally, CCMC will provide Continuous/Real-time run services for OSPREI, with results publicly available on CCMC's Integrated Space Weather Analysis System (iSWA, \url{https://ccmc.gsfc.nasa.gov/iswa/}). Real-time runs will use CME input parameters from CCMC's Space Weather Space Weather Database Of Notifications, Knowledge, Information (DONKI, \url{https://kauai.ccmc.gsfc.nasa.gov/DONKI/}). We note that real-time space weather activities that require human-in-the-loop analyses, previously performed by CCMC staff, have transitioned from CCMC to the Moon to Mars (M2M) Space Weather Office (\url{https://science.gsfc.nasa.gov/674/m2m/}). The M2M Space Weather Analysis Office was established in 2020 to support NASA Johnson Space Center (JSC) Space Radiation Analysis Group (SRAG) with human space exploration activities by providing expert analysis of the space radiation environment. NOAA Space Weather Prediction Center remains the primary forecasting operations office for SRAG. M2M provides secondary expert support for SRAG by analyzing state-of the-art space radiation models tailored to SRAG’s needs. M2M also supports NASA robotic missions with space weather notifications and anomaly analysis support. The M2M team populates CCMC’s DONKI and CME Scoreboard during their real-time analysis of space weather conditions, and sends certain real-time simulation results to CCMC's iSWA. CCMC continues to be the primary interface with model developers for all model onboarding, including real-time space weather models. CCMC is also continuing all other real-time space weather activities including: developing real-time systems, running real-time simulations, ingesting and serving information through CCMC’s iSWA, DONKI and Scoreboards.

\section{Observed CME}
To illustrate its capabilities, we apply OSPREI to two recent CMEs, the 2021 May 09 and the 2021 April 22 CME, hereafter CME 1 and CME 2 respectively. These CME are chosen for no particular reason other than they are a relatively recent CMEs that are more dynamic than merely a small blob convecting out with the background solar wind. As we will show, the CME 1 has a more prominent in situ structure than CME 2, despite relatively similar coronal properties. To mimic the approach that would be used for a real-time prediction we use as much information as possible from DONKI.

\subsection{CME 1 - 2021 May 09}
\begin{figure}
 \noindent\includegraphics[width=\textwidth]{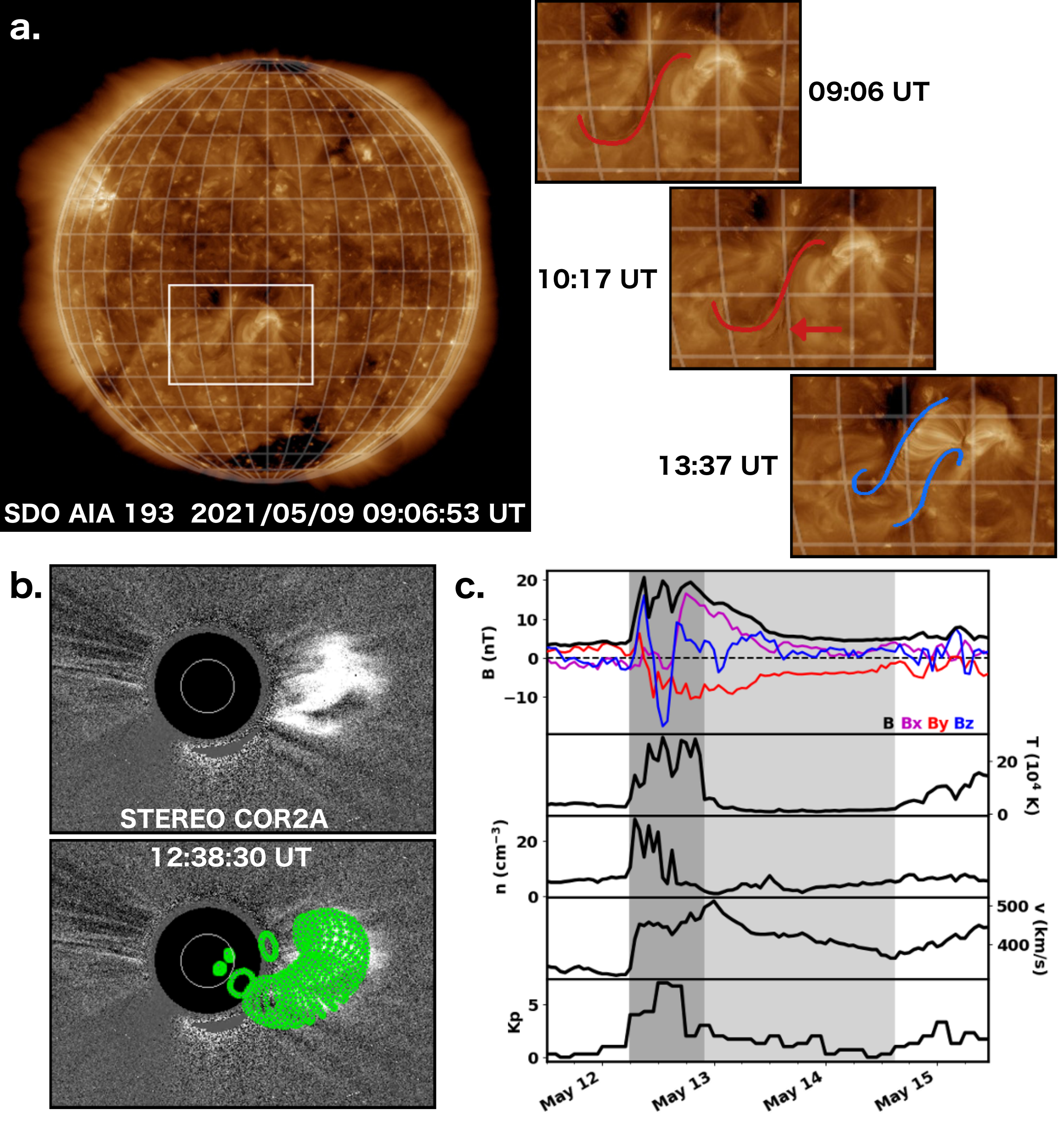}
\caption{Summary of observations for CME 1. (a)EUV image of the solar disk from SDO AIA 193 \AA{} at 09:06 UT on 2021 May 09 with the CME source region outlined and side panels highlighting it at 09:06 UT, 10:17 UT, and 13:37 UT. (b) Coronagraph images from STEREO COR2A at 12:38 UT showing the CME with and without the GCS reconstruction. (c) In situ observations from the OMNI database between 2021 April 24 and 27 around the time of CME arrival with the sheath (dark gray) and magnetic obstacle portions (light gray) highlighted.}
\label{obs1}
\end{figure}

DONKI lists this CME as first visible in STEREO COR2A at 11:23 UT on 2021 May 09, erupting from a location at S20E10 (Stonyhurst longitude, no specific active region listed, details at \url{https://kauai.ccmc.gsfc.nasa.gov/DONKI/view/CMEAnalysis/16870/1}). The source of this CME is a a filament eruption beginning at 10:06 UT, best visible in SDO AIA 193/304 \AA{} centered around S20E10, but stretching in longitude from around E20 to E00. The eruption is also visible in the southwest of STEREO A EUVI 195 \AA{} imagery. DONKI lists the CME as reconstructed in the corona at -4$^{\circ}$ latitude, -10$^{\circ}$ longitude, with a speed of 717 km/s and angular width of 27$^{\circ}$. We extend our observational analysis beyond the DONKI entry, but restrict our methods to those plausible for future forecasters on operational timescales. Figure \ref{obs1} presents a summary of the observational analysis.

We first look at the EUV signatures. Fig. \ref{obs1}(a) shows a full disk SDO AIA 193 \AA{} image from 09:06 UT with grid lines show 10$^{\circ}$ resolution in latitude and longitude. The source region is highlighted with a white box and the side panels show this region at three different times. A dark, S-shaped filament is visible, which we trace the bottom of in red in top side panel. The first noticeable motion occurs between 09:06 UT and 10:17 UT. The red outline is at the same position in the middle side panel and the arrow points towards small filamentary threads that have begun moving in the south. The southern portion of the filament begins propagating away by 10:30 UT, followed by the northern portion. By 12:00 UT, a post eruption arcade (PEA) extends the full latitudinal extent of the original filament. The bottom side panel shows the (PEA) at 13:37 UT with blue outlines showing the J-shape of its edges. Around 14:00 UT another CME erupts from the active region visible on the northeast of the disk but DONKI lists it as propagating at N12E60 with a speed of 580 km/s, so we do not expect any interaction between it and CME 1.

We fit a GCS model to coronagraph observations of CME 1, to compare with the DONKI results and to add a measure of the tilt. Fig. \ref{obs1}(b) shows STEREO COR2A results at 12:38:30 UT, both without and with Graduated Cylindrical Shell reconstruction \cite<GCS, >{The06}. We fit the position as S02E09 and the tilt at 45$^{\circ}$ at a height of 7.9 $R_S$.

DONKI associates CME 1 with an interplanetary shock at 05:48 UT at 2021-05-12 with a ``Quality 2'' interplanetary shock/CME arrival, which corresponds to a clear CME shock arrival signature. Fig. \ref{obs1}(c) shows hourly OMNI in situ data around this time span with the top panel showing magnetic field (black is total B, purple, red, and blue are $B_x$, $B_y$, and $B_z$ in GSE coordinates), the second panel showing the temperature, the third panel showing the number density, the fourth panel showing the velocity, and the bottom panel showing the Kp index. We show the data at relative coarse resolution as our model is not capable of producing any small scale variations. The shaded region corresponds to our CME boundaries with the dark gray in front designated as the sheath and the light gray the CME itself. While we see a clear magnetic field enhancement and some rotation, we refer to this section as the magnetic obstacle (MO) as done in \citeA{Nie18Wind} to avoid any worry whether it fully satisfies the definition of an in situ flux rope \cite{Bur81}. We note that FIDO results have shown that a sufficiently oblique and/or flank encounter with a CME defined from flux rope equations can produce an in situ profile without the canonical smooth rotation. We still refer to the part of the FIDO simulation that corresponds to the MO as a flux rope as it is a direct result of a flux rope equation. We set the boundary between the sheath and MO on 2021-05-12 at 22:00 UT based on the sudden change in temperature. We set the end of the MO at 2021-05-14 at 15:00 UT based on the velocity and temperature starting to increase. These two boundaries are estimates and one could potentially argue for different values but these regions are only used for illustrative purposes and not any quantitative analysis. To mimic a real time forecast we do not use any in situ information about the CME when setting the initial parameters as it would not be available. We do use the upstream in situ solar wind information, which is used for the interplanetary drag and expansion and the sheath calculation, as our focus is not on forecasting the solar wind itself and OSPREI can be coupled with any model designed for that.

\subsection{CME 2 - 2021 April 22}
\begin{figure}
 \noindent\includegraphics[width=\textwidth]{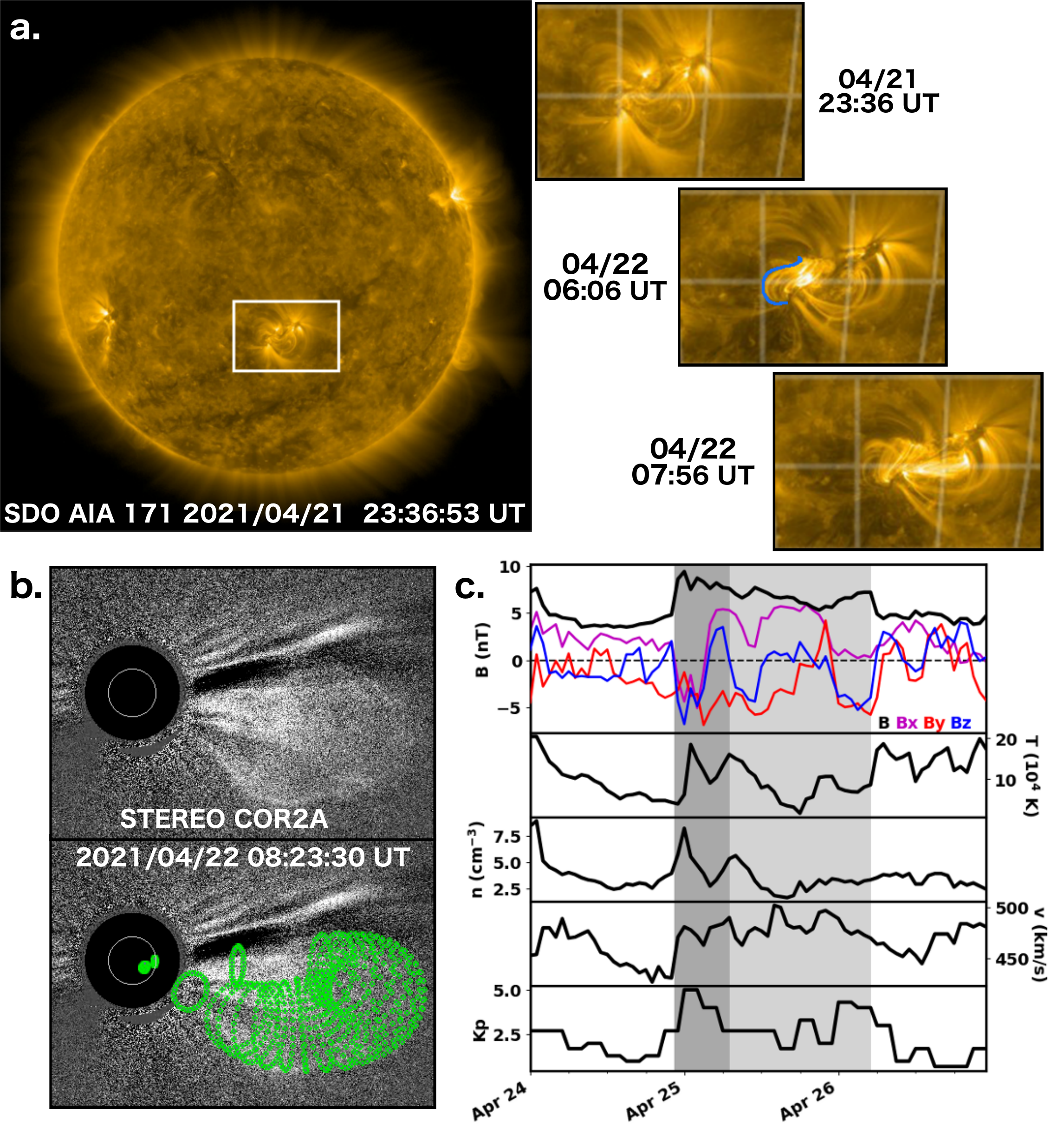}
\caption{(a) Composite EUV image of the solar disk from SDO AIA at 04:47 UT on 2021 April 22 highlighting the CME source region. (b) Coronagraph images from STEREO COR2A at 08:23 UT showing the CME with and without the GCS reconstruction. (c) In situ observations from the OMNI database between 2021 April 24 and 27 around the time of CME arrival with the sheath (dark gray) and magnetic obstacle portions (light gray) highlighted.}
\label{obs2}
\end{figure}

DONKI lists CME 2 as erupting from Active Region 12814 at S24W08 and is reconstructed in the corona at S13E08 with a reported speed of 803 km/s and angular width of 30$^{\circ}$ (\url{https://kauai.ccmc.gsfc.nasa.gov/DONKI/view/CMEAnalysis/16774/2}). Figure \ref{obs2} presents an observational summary of CME 2, analogous to Fig \ref{obs1} for CME 1.

We look at multiple EUV channels but only include the most relevant one, 171 \AA{}, in Fig. \ref{obs2}(a), which differs from showing 193 \AA{} in Fig. \ref{obs1}(a). Subtle motion in the active region loops can be seen as early as 23:36 UT on 2021 April 21. The loops in the eastern portion begin rapidly brightening around 04:30 UT, and by 07:30 UT the PEA extends to the western portion (see middle and bottom side panels in Fig. \ref{obs2}(a)). In the middle side panel, we outline in the edge of the loops to highlight their reverse-J shape.

Fig. \ref{obs2}(b) shows CME 2 in the corona, with and without our GCS reconstruction. By 08:23 UT, the CME has deflected northward to -10$^{\circ}$ latitude and eastward to -1$^{\circ}$ Stonyhurst longitude, with a tilt of 20$^{\circ}$ at a height of 14.4 $R_S$.

Fig. \ref{obs2}(c) shows the OMNI in situ data near the time of the CME arrival. The CME is associated in DONKI with a shock at 22:24 UT on April 24, which matches the sudden increases in the in situ parameters. We set the internal boundary between the sheath and MO at 07:00 UT on April 25 and the end boundary of the MO at 5:00 UT on April 26 based on sharp changes in the in situ measurements. We again note that these last two boundaries are estimates, especially for this CME where the in situ signatures are not that much stronger than the background solar wind itself.

\section {OSPREI Inputs}
We now address how we determine OSPREI inputs from the information that would be available for a real-time forecast. In this section we outline the general process for determining the inputs then describe the identification of inputs for both example CMEs. 

Table 1 lists all 34 parameters that can be varied for an OSPREI simulation. All three components of OSPREI are fully-determined by these numbers and the PFSS background driven by the synchronic magnetogram. The table shows the parameter with the variable name in parenthesis (as used in Fig. \ref{OSP}), the value used for our test cases, and the source for those values. If a parameter is varied within our ensemble then we also include the range considered. Parameters are grouped by OSPREI component (describing either the CME itself or the empirical coronal evolution models), background solar wind, or satellite properties. 

While at first glance this may appear an unwieldy number of parameters to set for predictions, we feel it can be managed and that it is the minimum necessary to describe a CME's full behavior from the Sun to the Earth. First, we emphasize that this is a completely explicit list of every free parameter that could potentially be changed. For our test case we leave 13 parameters at default values, as we expect would be done for actual forecasts. Of the 21 parameters remaining, 17 of them are known or can be reasonably estimated from various sources. The position and orbit of the Earth or satellite are known (Lat, Lon, Dist, Orbit) and we expect that any forecasting operation would have a real-time model of the background solar wind to provide those four inputs (velocity, density, magnetic field, and temperature). For our illustrative example cases we use OMNI values of the upstream solar wind. 

From EUV observations we can identify the start time, initial latitude, longitude, and tilt. We can also infer the flux rope's handedness from various EUV observables related to the filament or post eruption arcade \cite{Pal17}. Some techniques require more detailed analysis than may be possible for a forecaster on short timescales, but often obvious S- or J-shaped signatures may hint at the handedness. In other cases, the Bothmer-Schwenn relationship, which links the handedness to the hemisphere from which the CME erupts, can be used, but this is a statistical relationship that only holds true about 75\% of the time. The GCS measurements provide estimates of the CME's angular width and, using distances measurements from multiple times, the coronal velocity. Finally, combining information from the EUV imagery and GCS reconstructions (time of first EUV motion, start of dynamic EUV behavior, GCS timing) we can narrow down the empirical propagation model, though these values are not as well constrained as some of the other inputs. We assume that the rapid acceleration phase begins when the dynamic EUV activity begins, so the initial CME speed and distance of this transition can be adjusted so that the timing matches the observations. Typically, we leave the end of the rapid acceleration phase at 10 $R_S$ but if one has thorough velocity measurements over distance there may be a preferable distance.

This leaves 4 ``difficult'' parameters, or ones for which default values are not well-constrained. First is the perpendicular angular width of the CME, which can technically be estimated from the $\kappa$ parameter in the GCS fit but we find that this parameter is rarely accurately adjusted for scientific reconstructions, let alone real-time predictions, so this is essentially unconstrained beyond what seems like a ``reasonable'' guess. Similarly, the mass of a CME can be reconstructed from white-light images \cite<e.g.>{Plu19} but with large uncertainty \cite{Vou10,deK17} and this is not commonly done for reconstructions. We again pick a reasonable value, assuming mass tends to scale with CME speed and size.

The initial magnetic field strength of the flux rope and internal CME temperature are largely unconstrained. These are two critical parameters for determining the interplanetary behavior and not having a good measure of them for real-time predictions is a major limitation, as it will be for any model sophisticated enough to include forecasts of these internal properties. It is not to say that methods have not yet been developed that could provide these quantities, such as scaling to flare, active region, or dimming properties \cite{Qui07,Gop17, Dis19}. \citeA{Sco19} simulated two CMEs using the 3D MHD model EUHFORIA where the magnetic properties of the spheromak CME are initiated using the Flux Rope from Eruption Data model \cite<FRED,>{Gop17}, which uses the post eruption arcade to estimate the reconnected magnetic flux that goes into forming a CME.  We have not yet tested any of these models in OSPREI. This will be the focus of future work, determining which methods work best for OSPREI, in general, as well as which can reasonably be applied on the timescales needed for forecasts. Additionally, these values are specified for the beginning of the ANTEATR component, rather than at the solar surface so any values derived for the properties at the time of eruption must be evolved to their outer coronal values. We already include evolution of these properties via conservation laws within ANTEATR but we have not shown that these relations would produce appropriate behavior during the much more dynamic expansion in the low corona or if changes, such as varying the adiabatic index, would be required. For this work, we simply pick reasonable values for the magnetic and thermal scaling parameters, assuming that these values tend to increase with CME size/speed.

\subsection{CME 1}
We assign CME 1 a start time of 09:00 UT based on the earliest observed EUV activity associated with this eruption. DONKI lists the source region at -12$^{\circ}$ latitude and -10$^{\circ}$ Stonyhurst longitude (17.6$^{\circ}$ Carrington longitude). We confirm this with the EUV observations and also visually estimate an initial tilt of 45$^{\circ}$.  We choose an $AW$ of  30$^{\circ}$ and a coronal $v_F$ of 800 km/s, which are slightly rounded up from the DONKI values to allow for continued evolution to the end of the ForeCAT simulation at 10 $R_S$. We set the slow rise velocity at 40 km/s and the transition to rapid acceleration at 1.5 $R_S$ to match the observed increase in EUV activity around 12:00 UT. We assign right-handedness to the flux rope due to the presence of forward S- and J-shaped structures in the EUV images, which matches the expected value from the Bothmer-Schwenn relationship. The CME mass is set to 3$\times$10$^15$ g, $AW_{\perp}$ to 10$^{\circ}$, and $\beta_M$ and $\beta_T$ both to 4 as reasonable values but without any real observational justification.

Based on OMNI observations approximately 6 hours upstream of the CME-driven shock impact, we set the solar wind velocity at 350 km/s, the number density at 5 cm$^{-3}$, the magnetic field strength at 5 nT, and the temperature at 7$\times$10$^4$ K. The satellite position corresponds to that of Earth at the start time with the distance set to L1 to mimic a near-Earth satellite. All other input parameters are left at default values.

\begin{table}
 \caption{Full List of OSPREI Free Parameters}\label{table}
 \centering
 \begin{tabular}{l r r l}
 \hline
  Parameter  &  CME 1  & CME 2 & Source \\
 \hline
   \textbf{ForeCAT CME Properties} &   &  &   \\
    & 2021/05/09  & 2021/04/21  & EUV  \\
   Start time ($t_0$)  & 09:00 UT &  22:00 UT &   \\
   Starting nose height ($R_{F0}$) & 1.1 $R_s$ & 1.1 $R_s$ & \textit{Default}  \\
   Initial latitude (Lat) & -12$\pm$1$^{\circ}$ & -24$\pm$1$^{\circ}$ & EUV/DONKI  \\
   Initial longitude (Lon) & -10$\pm$1$^{\circ}$ & 1$\pm$1$^{\circ}$& EUV/DONKI  \\
   Initial tilt (Tilt) & 45$\pm$1$^{\circ}$ & 20$\pm$1$^{\circ}$  & EUV   \\
   Max coronal velocity ($v_F$) & 800$\pm$50 km/s & 800$\pm$50 km/s  & GCS/DONKI  \\  
   Max coronal angular width (AW) & 30$\pm$5$^{\circ}$  & 30$\pm$5$^{\circ}$   & GCS/DONKI  \\
   Max coronal perpendicular AW (AW$_{\perp}$) & 10$\pm$1$^{\circ}$   & 10$\pm$1$^{\circ}$  &  --  \\
   Maximum mass ($M_{CME}$) & 3$\pm$0.5$\times$10$^{15}$ g & 2$\pm$0.5$\times$10$^{15}$ g & --  \\
   Coronal axis aspect ratio ($\delta_{Ax}$)  & 0.75$\pm$0.1 & 0.75$\pm$0.1 & \textit{Default}  \\
   Coronal cross section aspect ratio ($\delta_{CS}$)  & 1$\pm$0.1 & 1$\pm$0.1 &  \textit{Default}  \\
 \hline
   \textbf{ForeCAT empirical models}  &  &  &   \\
   Initial slow rise velocity ($v_{F0}$)  & 40 km/s & 50 km/s &  EUV/GCS*  \\
   Start of rapid acceleration ($R_{v1}$) & 1.5 $R_s$ & 2.0 $R_s$  & EUV/GCS*  \\
   Height of max coronal velocity ($R_{v2}$) & 10 $R_s$ & 10 $R_s$ & \textit{Default}  \\
   Expansion model initial AW (AW$_0$)  & 5$^{\circ}$ &  5$^{\circ}$ & \textit{Default}  \\
   Expansion model length scale ($R_{AW}$)  & 1 $R_s$ & 1 $R_s$  & \textit{Default}  \\
   Height of max mass ($R_M$) & 10 $R_s$ & 10 $R_s$ & \textit{Default}  \\
   \hline
   \textbf{ANTEATR CME Properties}  &  &  &   \\
   Flux rope magnetic scaling ($\beta_M$) & 4$\pm$0.25 & 2$\pm$0.25 &  -- \\
   Elliptical flux rope model powers ([m,n]) & [0,1] & [0,1] &  \textit{Default} \\
   Elliptical flux rope model twist ($\tau$) & 1 & 1 &  \textit{Default} \\
   Elliptical flux rope model tor/pol ratio ($C$) & 1.927 & 1.927 &  \textit{Default} \\
   Flux rope temperature scaling ($\beta_T$) & 4$\pm$0.25 & 2$\pm$0.25 &  -- \\
   Adiabatic index ($\gamma$) & 1.33$\pm$0.1 & 1.33$\pm$0.1 &  \textit{Default} \\
   Interplanetary expansion factor ($f_{Exp}$) & 0.5$\pm$0.1 & 0.5$\pm$0.1 & \textit{Default} \\
 \hline
   \textbf{FIDO CME Properties}  &  &   \\
   Flux rope handedness (+/-) & Right  & Left & EUV \\
 \hline
   \textbf{Solar Wind Properties}  &  &  &   \\
   Solar wind 1 AU velocity & 350$\pm$25 km/s & 450$\pm$25 km/s & OMNI \\
   Solar wind 1 AU density & 5$\pm$1 cm$^{-3}$ & 5$\pm$1 cm$^{-3}$ & OMNI \\
   Solar wind 1 AU magnetic field & 5$\pm$1 nT & 4$\pm$1 nT &  OMNI \\
   Solar wind 1 AU temperature & 4$\times$10$^4$ K  & 7$\times$10$^4$ K & OMNI \\
   Drag coefficient & 1 & 1 &  \textit{Default} \\
   \hline
   \textbf{Satellite Parameters}  &  & &    \\
   Latitude  & -3.3$^{\circ}$ & -5.0$^{\circ}$ &  Earth location \\
   Longitude  & 0$^{\circ}$ (27.6$^{\circ}$) & 0$^{\circ}$ (257.5$^{\circ}$) &  Earth location  \\
   Distance  & 213 $R_s$  & 213 $R_s$  & L1  \\
   Orbital Speed & 2.8$\times$10$^{-6}$ rad/s & 2.8$\times$10$^{-6}$ rad/s & Earth's orbit  \\
 \hline
\multicolumn{4}{l}{All longitudes in Stonyhurst coordinates at the start time}
\end{tabular} 
\end{table}

\subsection{CME 2}
We follow the same process for CME 2, taking the initial timing and location from the EUV images and DONKI entry. We estimate a less inclined tilt of 20$^{\circ}$ for this CME. DONKI has similar values to CME 1 for the angular width and the speed of this CME so we again use 30$^{\circ}$ and 800 km/s. The EUV images show a longer delay between the beginning of any activity and the explosive, impulsive activity so we move $R_{v1}$ to 2 $R_S$, allowing CME 2 to spend more time in the slow rise phase. In determining the handedness, the EUV signatures are not quite as distinct as for CME 1, but are most consistent with a reverse J shape, implying a negative handedness, opposite as expected from the Bothmer-Schwenn relationship. This CME appears less intense than CME 1 in the coronagraph images so we decrease the CME mass to 2$\times$10$^{15}$ g and $\beta_M$ and $\beta_T$ both to 2. The solar wind values are again taken from OMNI and the satellite position determined by the Earth's location at the time of eruption.

\section{OSPREI Results}
The following subsections show our OSPREI results for these CMEs. We emphasize that our focus is on presenting OSPREI as a tool used in a real-time fashion and the outputs available for predictions, not the details of this specific CME. We have not fine-tuned any parameters to achieve a perfect fit. It is possible that the results could be improved for a detailed scientific study of this event, but our ``forecast'' approach still yields very encouraging results.

We run an ensemble of 100 members and allow for variations in 17 different initial parameters, as shown in Table \ref{table} . The range of the ensemble variations is meant to represent the uncertainty in reconstructed values or our best guesses for plausible values. The full ensemble requires about an hour to run on an average laptop. Any forecasting center would likely have computers with more sophisticated processing power, potentially reducing the run time to tens of minutes.

\subsection{Coronal Predictions}
\begin{figure}
 \noindent\includegraphics[width=\textwidth]{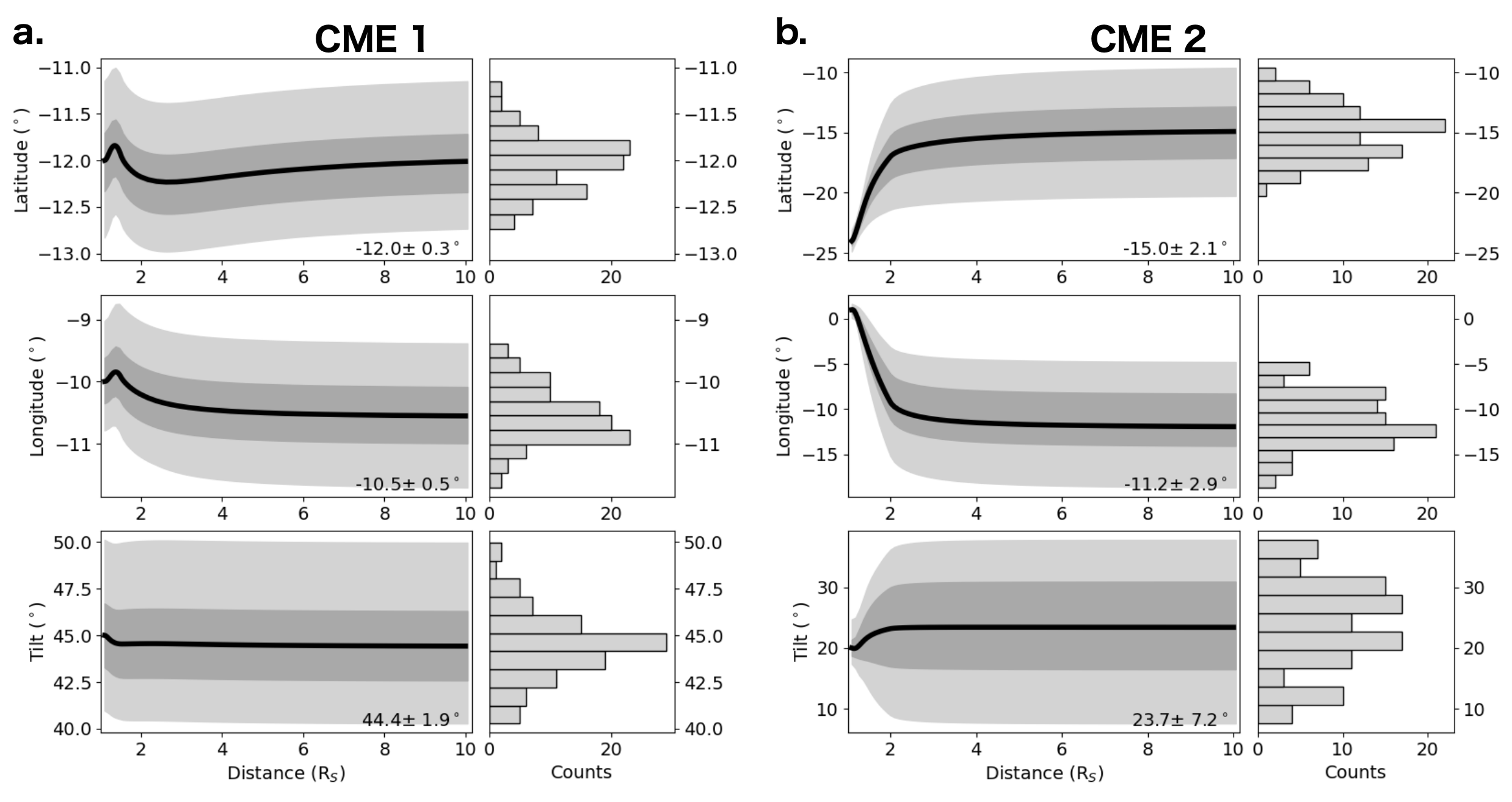}
\caption{ForeCAT results for CME 1 (left) and CME 2 (right) showing the latitude (top), longitude (middle), and orientation (bottom) versus distance. Each panel includes the ensemble seed profile (black), the full range of the ensemble (light gray), and one standard deviation about the mean (dark gray). The histograms on the right show the distribution of the final values for each parameter.}
\label{cpa}
\end{figure}

We begin with the coronal predictions. Figure \ref{cpa} shows the change in latitude (top panel), longitude (middle), and tilt (bottom) versus distance from ForeCAT, with results for CME 1 on the left and CME 2 on the right. For each CME, the left panels show the coronal trajectory The black line shows the ensemble seed, the dark gray shows the core of the ensemble (one standard deviation about the mean), and the light gray shows the full range. The numbers in the bottom right display the mean and standard deviation for each value at the end of the ForeCAT simulation. The histograms in the right panels show the distribution of the ensemble values at the end of this component.

Fig. \ref{cpa}(a) shows a negligible amount of coronal deflection or rotation for CME 1. While we see a small imbalance in the forces causes a minor northwestern deflection but this immediately reverses to the southeast and stabilizes at values nearly identical to the initial ones. The behavior is nearly identical for all ensemble members with the profiles simply being shifted up or down according to their initial values. The mean final coronal latitude, longitude, and tilt are all within 0.5$^{\circ}$ of the ensemble seed values. The ranges in the final ensemble latitude and longitude are actually smaller than the range in the initial values of those parameters. 

Fig. \ref{cpa}(b) shows the expected northeastern deflection, as observed for CME 2, for the majority of ensemble members. On average, these CMEs deflect 9$^{\circ}$ northward and 12$^{\circ}$ eastward, but have negligible rotation (4$^{\circ}$ counterclockwise). The full ensemble range exhibits some significant rotations in both directions but the average behavior is small. We typically interpret this spread in rotations as consistent with a background magnetic field causing little to no rotations and that they instead result from slight imbalances in our chosen initial position. 

These two exhibit fairly different behavior, a nearly radial trajectory compared to about 10$^{\circ}$ of change in both latitude and longitude. For both CMEs, the ForeCAT results yield a position and orientation consistent with the DONKI and GCS reconstructions but additionally provide a range of probabilities based on a physics-driven simulation.

\subsection{Interplanetary Predictions}
\begin{figure}
 \noindent\includegraphics[width=0.75\textwidth]{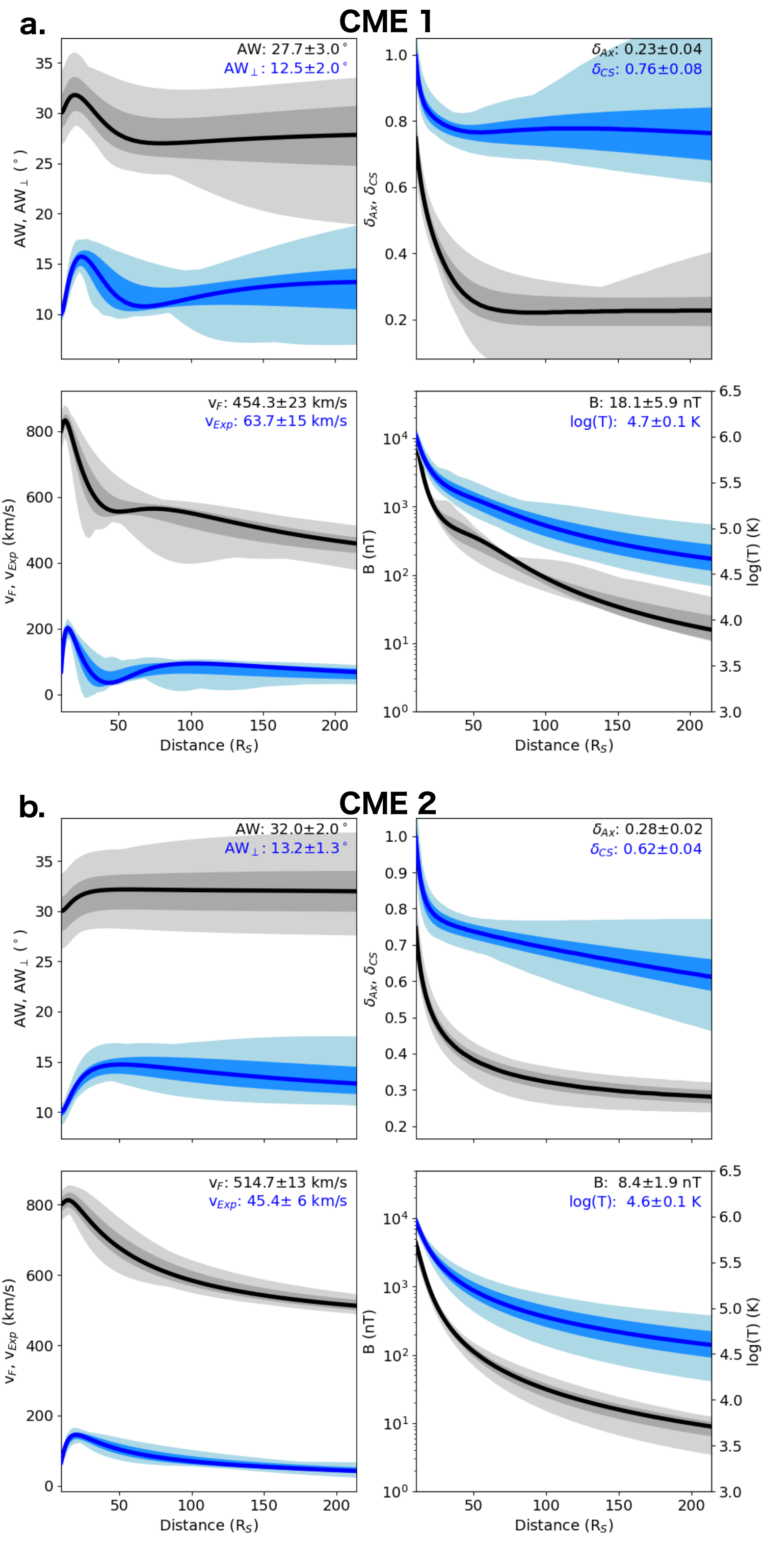}
\caption{Interplanetary evolution for CME 1 (left) and CME 2 (right) in the same format as Fig. \ref{cpa}. Within each subfigure, the top left shows the angular width (black) and perpendicular angular width (blue). The top right shows the axial (black) and cross section (blue) aspect ratios. The bottom left shows the front velocity (black) and the expansion velocity (blue). The bottom right shows the magnetic field strength (black) and temperature (blue).}
\label{drag}
\end{figure}

Next we consider results for the ANTEATR interplanetary predictions. Figure \ref{drag} shows the evolution of parameters at interplanetary distances, the shading is a similar format to Fig. \ref{cpa} but all panels include a second parameter in blue. ANTEATR does generate additional results not shown here, such as the speed of the central axis in both the radial and perpendicular directions, but we limit this figure to the values most relevant to provide a clear picture for forecasting. The mean value and standard deviation at the end of the simulation are displayed for each parameter.

The top left panels show the evolution of the angular widths, $AW$ and $AW_{\perp}$. CME 1, which has stronger internal magnetic field and higher temperature than CME 2, shows a rapid initial expansion in both $AW$ and $AW_{\perp}$ until 20 $R_S$  then begins contracting. $AW$ remains relatively stable beyond 50 $R_S$ but $AW_{\perp}$ continues slowly increasing out to 1 au. CME 1's behavior results from the initial magnetic and thermal overpressure causing expansion until the CME is under-pressured, and eventually reaches an equilibrium state. In comparison, CME 2 shows a more gradual initial increase in the angular widths with $AW$ reaching equilibrium and $AW_{\perp}$ slowly contracting beyond 50 $R_S$. We see extreme behavior from a few ensemble member, particular for CME 1 with the stronger magnetic and thermal forces. These cases rapidly expand then rapidly contract, ultimately arriving at 1 au with a smaller $AW$. For $AW_{\perp}$, the rapid contraction can be followed by more expansion if the CME becomes overpressured relative to the background again. Whether or not these extreme cases alternating between expansion and contraction are realistic is open for debate but we note that the changes typically remain within $\pm$5$^{\circ}$ of the initial values. In contrast, the core of both ensembles mirrors the seed case profile with slight changes in the parameter values. 

The top right shows the change in the axis and cross section aspect ratios, $\delta_{Ax}$ and $\delta_{CS}$. Both parameters decrease for both CMEs, which corresponds to a pancaking effect where the radial direction becomes relatively shortened due to the combined effects of drag and axial magnetic forces. Interestingly, we see a slightly larger change in $\delta_{Ax}$ for CME 1 than CME 2 but a larger change in $\delta_{CS}$ for CME 2. CME 1 has a more significant initial decrease in $\delta_{Ax}$ but then reaches a nearly constant value by 75 $R_S$. In comparison, $\delta_{Ax}$ initially decreases more slowly for CME 2, but a gradual decrease continues until 1 au. For $\delta_{CS}$, CME 2 exhbits a slightly more rapid initial decrease which transitions to a more gradual but continued decrease out to 1 au. CME 1 has $\delta_{CS}$ decrease slower, then remain nearly constant beyond 40 $R_S$. There is a slight increase then decrease beyond this distance but of insignificant magnitude. Again, the majority of the ensemble behaves like the seed profiles with small changes in magnitude and we see a few extreme cases, which may have borderline unstable inputs.

The bottom left panels shows the speed of the CME front, $v_F$, and the expansion speed, $v_{Exp}$. Technically, we show the rate at which the radial cross section length, $r_r$ in Fig. \ref{OSP}, increases, but this is analogous to the in situ expansion speed that would be observed for a direct hit at the CME nose. For both CMEs, $v_{Exp}$ shows an initial increase due to the overpressure relative to the background, which causes the increase in $AW_{\perp}$, but then decreases as the external pressure becomes larger than that inside. CME 2 gradually decreases over time whereas CME 1 exhibits times of acceleration and deceleration out to about 100 $R_S$, after which $v_{Exp}$ slowly decreases $v_F$ combines the effects of the cross-sectional expansion with the axial expansion and the bulk CME propagation. For both CMEs, $v_F$ shows an initial increase, which is driven by the expansion of the cross section pushing the nose forward, but quickly begins decreasing as both the drag and axial magnetic forces act to slow down the forward motion. We see the same acceleration and deceleration in the $v_F$ profile for CME 1, this is a direct result of $v_{Exp}$, not any additional forces on the axial expansion or bulk motion.

The bottom right shows the internal magnetic field strength (taken at the central axis) and the logarithm of the temperature. Both parameters decrease as the CME expands, as expected, but this gives a most probable value and the range for these internal properties upon impact, which is uncommon for a simple analytic prediction.

\begin{figure}
 \noindent\includegraphics[width=\textwidth]{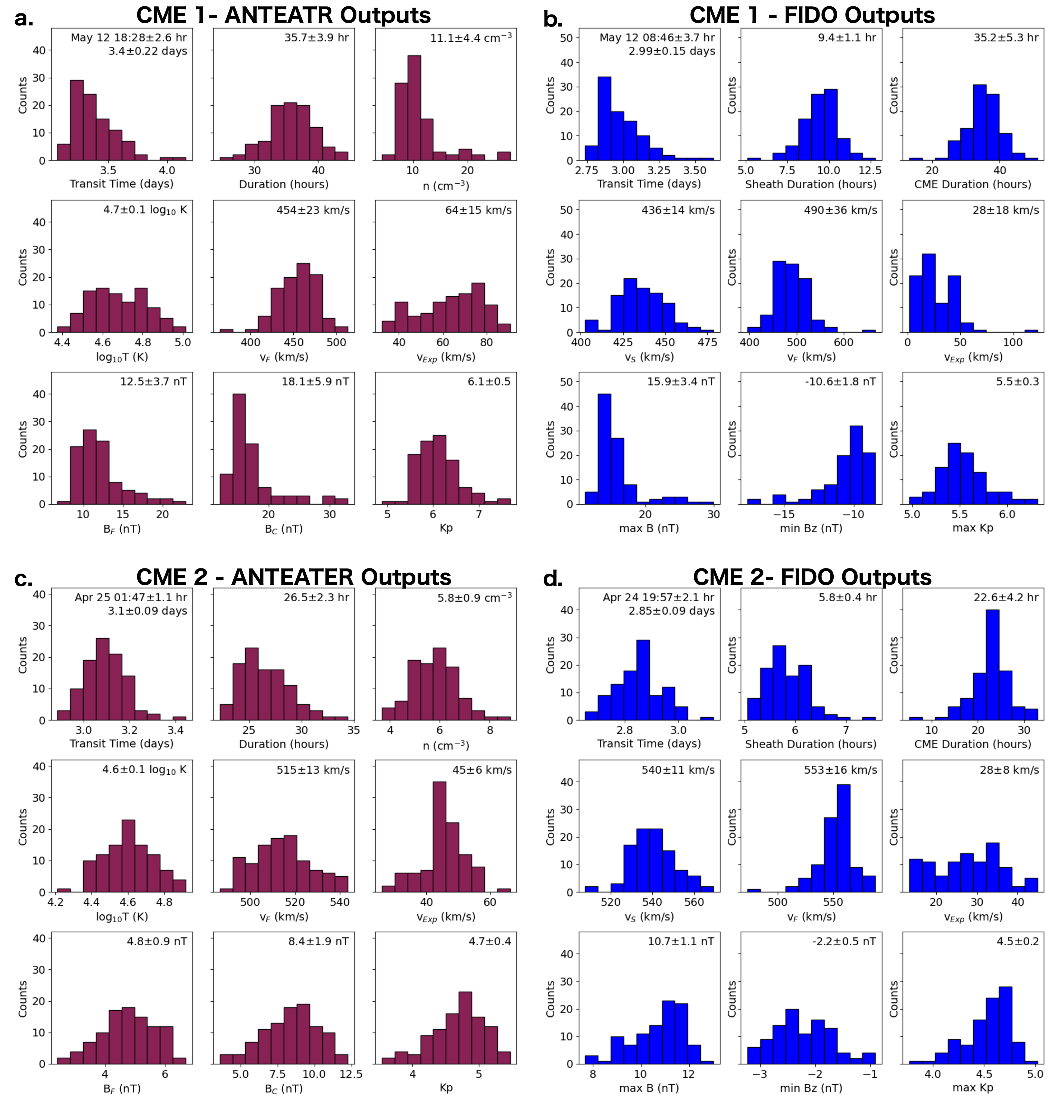}
\caption{Histograms of the outputs for CME 1 from ANTEATR (a) and FIDO (b) for CME 2 from ANTEATR (c) and FIDO (d). Where possible, comparable values between ANTEATR and FIDO are presented in analogous panels (i.e. the front velocity in the center panels). }
\label{hist}
\end{figure}

Instead of looking at full profiles, forecasters may prefer to simply see the range of possibilities at the time of impact. Figure \ref{hist}(a) shows histograms of different values from ANTEATR. ANTEATR does not simulate the full path of the satellite through the CME, that is only performed in FIDO, so properties such as the duration are estimated from the geometry. The histograms include $v_F$ and $v_{Exp}$ as in Fig. \ref{drag} and are global properties of the CME rather than what one may observe in situ for that specific impact location. As such, if available, outputs from FIDO (shown in Fig. \ref{hist}(b) and (c) and discussed in Section \ref{ISP}) should be prioritized over analogous ANTEATR outputs.

The top left panels shows the transit time. ANTEATR only simulates the flux rope, the sheath is not included until FIDO, so this corresponds to the arrival of the front of the MO. For CME 1 we find an average arrival time of 2021 May 12 at 18:28 UT (compared to the observed 22:00 UT) and 2021 April 25 at 01:47 UT for CME 2 (observed at 07:00 UT). The top middle panels shows an average durations of 36 hours and 27 hours for CME 1 and 2, respectively. This is about 5 hours short of the observed duration for CME 1 and 5 hours long for CME 2. This error is larger than the standard deviation of the ensemble values for both CMEs, but we note that the MO start and stop time used to derive the observed durations were only estimates. 

For CME 1 we find an average density of 11.1$\pm$4.4cm$^{-3}$ and temperature of $\log(T)=$4.4$\pm$0.1K (top right and middle left panels) and average values of 5.8$\pm$0.9cm$^{-3}$ and $\log(T)=$4.7$\pm$0.1K. The relative variation in the final density and temperature between the two cases is less than the variation in their initial properties. We find average propagation and expansion speeds of 450 km/s and 65 km/s for CME 1 and 515 km/s and 45 km/s for CME 2. However, CME 1 has stronger magnetic field than CME 2 (a $B_F$ of 13 nT versus 5 nT at the front and a $B_C$ of 18 nT versus 8 nT at the center). The background solar wind is much slower for CME 1 so the continued effects of drag cause the bulk speed to decrease but we still see more expansion due to the stronger initial internal magnetic field and temperature. The bottom left panels show the $Kp$ index, estimated at the front of the CME using $v_F$ and $B_F$.

\subsection{In Situ Predictions}\label{ISP}
\begin{figure}
 \noindent\includegraphics[width=\textwidth]{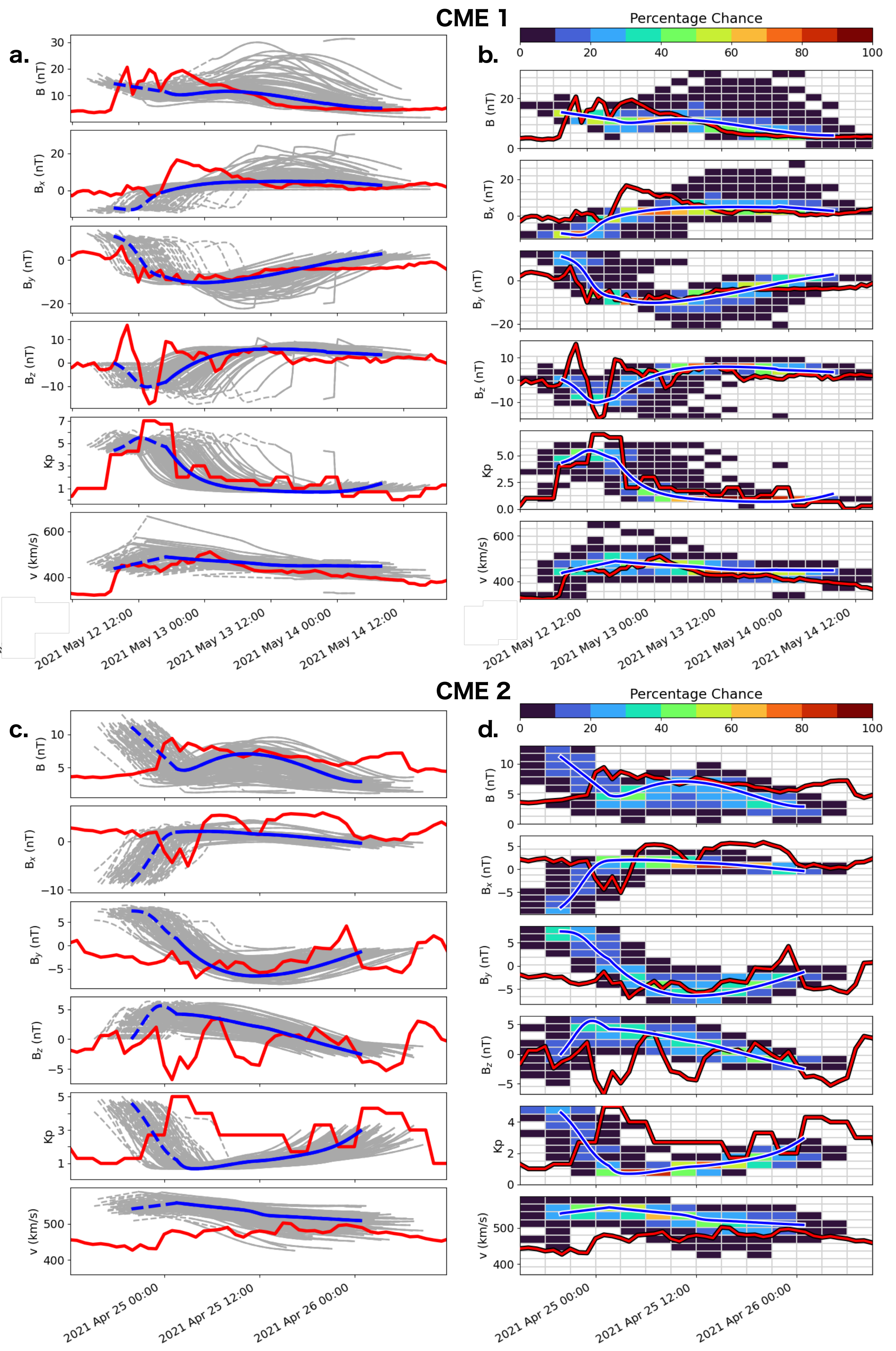}
\caption{(a) Individual in situ profiles for each ensemble member (gray) with the ensemble seed (blue) and OMNI observations (red). (b) Contours of the probability of specific in situ values over time derived from the density of the individual profiles. }
\label{IS}
\end{figure}

The next figures correspond to FIDO results for the synthetic in situ magnetic and velocity profiles. Figure \ref{IS} shows the full in situ profiles in two different formats for both CMEs. These figures provide more details than the histograms in Fig. \ref{hist}(b) and (d), which we discuss next, but may not be as easy to quickly interpret. From top to botom, each subfigure shows the total magnetic field strength ($B$), three components in GSE coordinates ($B_x$, $B_y$, $B_z$), the estimated Kp, and the in situ velocity profiles, from top to bottom. he observed in situ profiles are shown in red and the seed profile in blue. Figure \ref{IS}(a) and (c) shows the individual profiles for each ensemble member (gray line) with the sheath component of each member shown as a dashed line. Figure \ref{IS} (b) and (d) show the percentage chance for different values of each parameter. The values are weighted by the total number of CMEs that impact the spacecraft so that in many time cells the total percentage of any value is less than 100 as not all CMEs were present at that time. The top row corresponds to results for CME 1 and the bottom row to CME 2.

For CME 1, we find that the majority of the profiles are in general agreement with the observations, though we do see some strange outliers, which correspond to the irregular cases seen in Fig. \ref{drag}. In general, the gray lines tend to fall around the red observations. The model flux rope has an asymmetric B profile due to the inclusion of expansion, but we do no capture some of the strong enhancement at the beginning of the observed MO. Looking at the individual components, we match $B_y$ and $B_z$ for the majority the duration, and the difference in total magnitude is largely driven by an enhancement in $B_x$ in the first half of the MO. This enhancement is likely beyond the capabilities of an idealized flux rope model, such as used in FIDO. 

The shock model gives some useful information, but is by far the least developed component of the OSPREI chain. The duration is similar to the observed value and the total magnetic field strength is close to but slightly less than the observed value at the shock. Predicting the upstream orientation is difficult and currently limited to the Parker Spiral, which is then scaled according to the Rankine-Hugoniot laws, then connected to the front of the flux rope, giving a very idealized profile with no stochastic fluctuations. The sheath $B_y$ matches well to the observations and $B_z$ is reasonable but cannot capture the large fluctuations, but $B_x$ does not match. 

Overall the full profile matches the observed $Kp$ pretty closely with just a slight overestimate at the shock and slight underestimate where $B_z$ fluctuates into negative values. The flux rope portion is slightly low but close to the observed values. We also find good agreement with the in situ velocity, maintaining a close match until the back portion where the simulated CME tend to have higher speeds.

Fig.\ref{IS}(a) shows individual lines extending over a wide range in all parameters but by looking at the percentages in (b) we find that the majority of the ensemble behaves pretty similarly to the seed. These more brightly-colored, higher-percentage cells all fall relatively close to the red line of the observations.

Fig. \ref{IS}(c) and (d) show results for CME 2. Again, the seed profile and many ensemble members resemble the observations, in general. The comparison is certainly less favorable than for CME 1, which may be a result of this MO having a significanly less flux-rope-like profile. The seed CME duration is a bit short and the arrival a bit early but we see cases that extend the full observed duration. The observed $B_x$ and $B_z$ show large fluctuations that we would not expect FIDO to reproduce with a simple flux rope model. The modeled $B_x$ tends to be slightly closer to zero than the average of the observed fluctuations whereas the modeled $B_z$ has slightly stronger field. The fit to the $B_y$ component is acceptable, this is also the only component that shows any sort of consistent large scale rotation. For the sheath, the durations tend to match the observed value, but are shifted forward in time due to the modeled flux ropes arriving slightly later than the observed MO. The predicted magnetic field strength at the shock is similar but slightly stronger than the observed value. The modeled velocity tends to be a bit high over the full CME duration, which makes sense given that the transit time is a bit short and arrival occurs early. This may be improved by changing the interplanetary expansion factor, $f$, from its default value. This determines whether the CME is initially expanding self-similarly or convecting with the solar wind before the ANTEATR forces take over. Again, our focus is presenting the OSPREI tool and the outputs that can be generated for forecasts so we leave values at reasonable defaults that would be used in real-time predictions rather than fine-tuning them for a best fit.

At first glance, the Kp profile seems a particularly poor match but this is largely a timing issue. The maximum Kp, which occurs at the shock, matches the peak in the observed values. We also see an increae in Kp toward the back of the flux rope, which matches an increase in Kp at the back of the MO.

These figures contain many details that may not be necessary for routine forecasting, but the allow us to see the full profiles instead of reducing time-variations to single values. This can show important factors such as CME 1 only briefly having a value near the most negative $B_z$, as opposed to a extended duration due to the rotation in the flux rope magnetic field. They also allow us to examine individual cases and better understand some of the outliers, such as for CME 1, the very short case with high $B$ and low $v$ or the late arrivals with sharp jumps in $B$, which occur when the impact occurs at angle that causes the satellite to exit through the flanks. While the variation for CME 2 is less extreme, we still find cases that have shorter duration and very weak magnetic field strength.

As an alternative to the more detailed profiles, Fig. \ref{hist}(b) and (d) show histograms of various parameters, analogous to the ANTEATR results in Fig. \ref{hist}(a) but for a more accurate simulation of the CME-satellite interaction and including the CME-driven sheath. Where possible, we include analogous properties in the same location in both figures, but not all panels are directly comparable.

The top row shows the transit time, sheath duration, and CME duration (left to right). For both CMEs, the mean predicted shock arrival is within 3 hours of the observed value with CME 1 arriving late and CME 2 arriving early. The mean CME duration is the same as seen in ANTEATR for CME 1, suggesting our geometric estimate is fairly accurate, but we do see a slightly larger range from the ensemble members that interact at more oblique angles. For CME 2, the ANTEATR duration differs by 4 hours. This impact occurs closer to the edge of the cross section and we have a wider range of CME positions and orientations, many of which are more oblique impacts. CME 1 and 2 have a sheath durations of 9 hours and 6 hours but sheath speeds (middle left panel) of 440 km/s and 540 km/s. The speed differential between CME and the background solar wind is nearly identical for these cases so the difference in sheath properties must results from the other solar wind properties. 

The middle rows show the speed at the front of the flux rope (middle column) and the expansion speed (right). The expansion speed is calculated as is done for observations as one half of the difference between the speed at front and end of the flux rope portion. For both CMEs, these histograms show a higher front speed (490 km/s and 550 km/s) but lower expansion speed (28 km/s for both) for the in situ values at the actual impact location as compared to the ANTEATR values given for the nose.

The bottom rows shows the maximum magnetic field (16$\pm$3 nT and 11$\pm$1 nT) and the most negative $B_z$ (-11$\pm$2 nT and -2$\pm$1 nT) within each ensemble profile (left and center panels). The bottom right panel shows the maximum Kp, which is largest value calculated over the full profile from the velocity and magnetic field. For both cases we find that the ANTEATR value overestimated the maximum $Kp$ determined by FIDO since we did not have the exact magnetic field orientation. In addition to overestimating the flux rope $Kp$, we find that the largest values occurred in the sheath for both cases, highlighting the importance of using the FIDO results derived from the full in situ profiles.

\section{Further Analysis}
\subsection{Spatial Variations}
\begin{figure}
 \noindent\includegraphics[width=\textwidth]{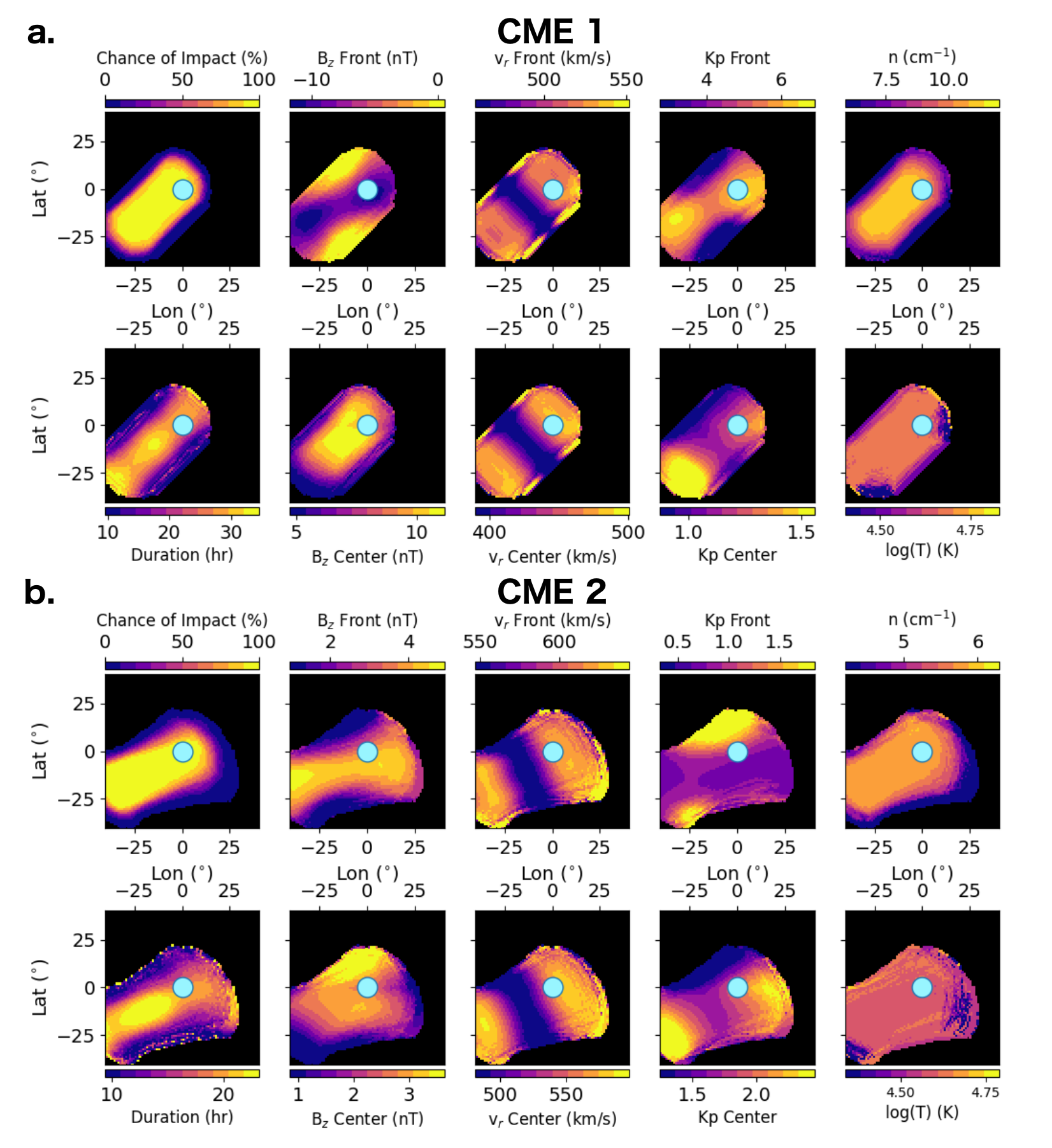}
\caption{Spatial contours of different values using the location and internal properties of the individual ensemble members for CME 1 (top) and CME 2 (bottom). The pale blue dot indicates the location of the Earth at the time of impact.}
\label{con}
\end{figure}

FIDO yields the full in situ profile for a specific trajectory through the CMEs, but it may be of use to see how things would change if the impact occurred at a slightly different location. We use the CME geometries and internal properties at the time of impact (from ANTEATR) to produce contour maps of various properties, shown in Figure \ref{con}. Each CME is projected onto a map of latitude and longitude with the Earth (or satellite) located at the origin (marked with a pale blue dot). For a single CME, this produces a rounded rectangular shape (see side and cross section views of the CME shape in the bottom of Fig. \ref{OSP}). We combine the projections of all ensemble members into a single map. For a given latitude and longitude in the map, we average the properties of all CMEs that project onto that point. This defines an area where there is a nonzero chance of CME impact. Near the edges of this area the properties may only be determined from only a few CMEs, if that is all that impacts that location, but toward the center the properties will tend to be the average of all CMEs.

The top left panels shows the percentage chance of impact. This is simply the sum of CMEs that impact that location divided by the total number of CMEs that reach the satellite distance. The denominator is not necessarily the same as the number in the ensemble, it is possible that the random variations lead to input parameters than make the internal magnetic forces unstable (e.g. combination of low mass with high magnetic field scaling and large $AW$). We do not include these ``failed'' unstable CMEs in the percentage calculation but we do include the misses that propagate the full distance and do not make contact. For CME 1 there is little variation in the position and orientation of the ensemble members so the contours exhibit the rounded rectangular shape that a single case would have and the percentages symmetrically decrease toward the edges due to slight variations in size. For CME 2, we see a wider range in positions and orientations, causing the area of non-zero contours to distort from the rounded rectangle. There is a large region with 100\% chance of impact but we see that the gradient in percentage change extends much farther to the right of it than it does to the left.

The bottom left panel shows the CME duration using the geometric estimate for the flux rope duration from ANTEATR, rather than a full in situ simulation value including the sheath, as from FIDO. Each location shows the average of all CMEs that impact there. For CME 1, the duration remains relatively constant as one moves along the toroidal axis (latitude and longitude both increasing or decreasing) but significantly decreases as one moves toward the edge of the cross section. The changes are similar for CME 2, but we see more variations near the edges of the nonzero region where the duration is derived from only a few CMEs.

The middle three panels on both rows show properties at the front and center of the CME (top and bottom respectively). From left to right we show $B_z$, the radial velocity $v_r$, and the estimated Kp. Both the RTN and GSE coordinate systems depend on the latitude and longitude of the observer. The values we show correspond to rotating these systems to each specific location within the figure, rather than using fixed systems at the position indicated by the pale blue dot. We note that we do include an estimate of the decrease in $B$ due to CME expansion and flux conservation during the time between the satellite interacting with the front and center of the CME. CME 1 is right-handed and CME 2 is left-handed but both have axial field pointing toward positive longitude. CME 1 has the most negative $B_z$ at the front, the exact pattern is determined by the axis and cross section geometry more so than any changes in field strength between different ensemble members. CME 2 does not have any negative $B_z$ at the front or center so the darkest regions correspond to the weakest field.

The velocity contours combine the effects of different parts of the axis and cross section expanding at different rates with the local orientation of the radial vector. With a pancaked CME the edges tend to be moving faster but at a more oblique angle to the radial. For both CMEs, this yields a faster radial velocity near the edges than at the nose. We also consistently find a larger speed at the front than at the center as both cases are still expanding at the time of impact.

The Kp contours combine the trends of the $B_z$ and $v_r$ panels. For CME 1, the impact happens to occur at one of the regions that combines the most negative $B_z$ and the fastest speed. Any change in the impact location would cause Kp to decrease, with the most rapid changes occurring toward the edges of the cross section. The Kp at the center is much weaker, as expected given the $B_z$ and $v$ maps. For CME 2, we see the strongest Kp at the front occurs just north of the impact location, but this is still only a Kp of 2. The Kp at the center is still weak but the strongest values occur near the flanks.

The right panels show the number density (top) and logarithm of the temperature (bottom). Both of these parameters are treated as uniform within a individual simulation so the spatial variations only result from the the actual spatial distribution of the ensemble CMEs, not any internal structure. We do not see as much variation in these parameters, with the exception of smaller density toward the edges of the nonzero area for CME 1 as this region corresponds to larger CMEs that have underwent more expansion.

\subsection{Analyzing Trends}\label{trends}
\begin{figure}
 \noindent\includegraphics[width=\textwidth]{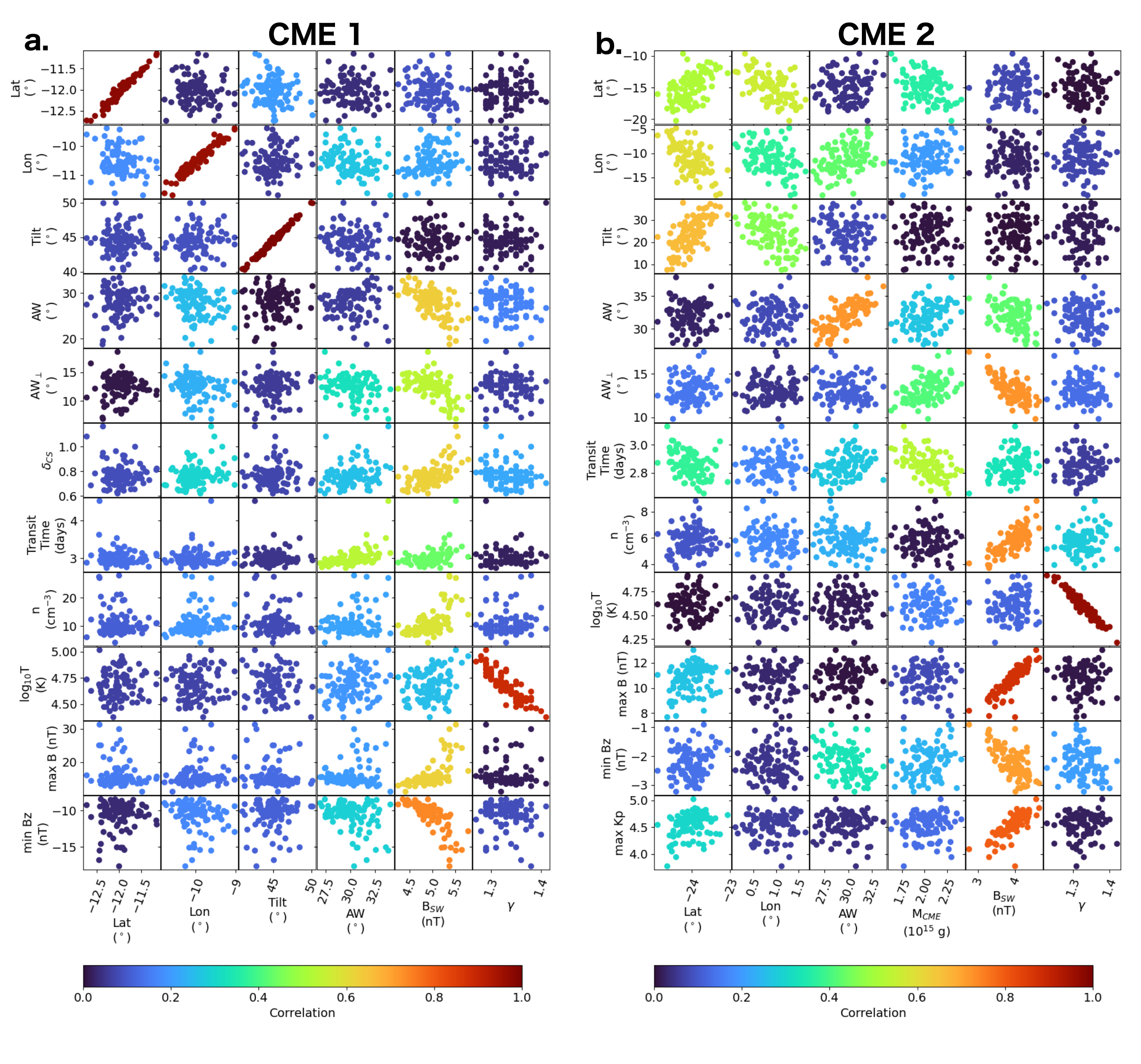}
\caption{Scatter plots showing the relation between variations in input parameters (horizontal axes) and outputs (vertical axes). The data in each panel is colored according to the correlation between that input-output pair.}
\label{corr}
\end{figure}

While the previous figures have shown the range of possibilities suggested by the ensemble modeling, they have not provided much, if any, reason for those variations. Figure \ref{corr} provides that information for when a forecaster wishes to dive deeper. The horizontal axis shows some of the inputs varied within the ensemble and the vertical axis shows significant outputs that can be quantified by a single number, such as those shown in the histograms. Each panel shows a scatter plot between an input and an output and are colored according to their correlation coefficient to help illuminate any trends. To avoid having a massive grid of panels, we set a minimum correlation and only include inputs and outputs that produces at least one correlation above this cutoff. For Fig. \ref{corr} we have chosen a cutoff of 0.5, but a user can vary this as suits their needs. 

For both CMEs we find 6 inputs and 11 outputs that exceed the correlation cutoff, but it is not entirely the same parameters for each CME. Some trends are expected, such as those between parameters that are both inputs and outputs. For CME 1, we saw little deflection and rotation so the final latitude, longitude, and tilt are extremely correlated with their input values. For CME 2, we only the final latitude is only strongly correlated to its initial value. Others correlation are less direct but still expected (e.g. higher $B_{SW}$ leading to more negative $B_z$ when the $B$ scaling is constant). In some cases we find that expected trends may be washed out by the model physics (e.g. final T is only slightly correlated with initial temperature scaling but is strongly correlated with the adiabatic index for both CMEs). Other times, unexpected trends may occur such as a correlation between $B_{SW}$ and the final $\delta_{CS}$ for CME 1, and a correlation between the initial latitude and final tilt of CME 2.

This figure also helps investigate outliers, such as seen in the in situ figures. The very short duration case in CME 1 corresponds to the lone dot with long transit time in Fig. \ref{corr}(a), which we can confirm with the full version of this figure without the correlation cutoff (not shown). This ensemble member happens to have a relatively low latitude, flat tilt, large $AW$, strong $B_{SW}$, and lower $\gamma$. This combination leads to relatively strong forces that are borderline unstable. For CME 2, we see fewer extreme outliers but can still identify cases such as the one with very low initial $B_{SW}$ that results in a short transit time and low density. While not exhibited by either of the test cases, it is possible that the ensemble results could cluster into different populations. For example, if the initial source region was similar to that of CME 2, so that we observed rotations in both directions, but with stronger forces we may have two populations corresponding to the clockwise and counterclockwise rotating groups. If the rotation is significant these populations may have very different magnetic field orientations, or one group may even correspond to non-impacts. While this level of interpretation may not be necessary for routine forecasting, it is essential to recognize when the majority of the ensemble falls within one group, making that the most probable forecast, there exists a small possibility of the observed behavior mirroring that of a subset instead of the mean values.

\section{Discussion}
In this work we have mimicked the process a forecaster would take, using only information available near the time of eruption (and OMNI data in place of a solar wind forecast) for two recent CMEs. This allows us to show the quality of results a forecaster could ideally achieve with OSPREI, as well as demonstrating the automatically-generated visualizations. While OSPREI presents a major step forward for forecasting CME-driven space weather effects in the near future, there are several steps that must be taken to further establish the model. 

First, more cases should be considered. We present two cases here that represent relatively similar coronal origins but fairly different in situ properties. OSPREI is able to reproduce the general behavior of both cases, but certainly better reproduces the in situ magnetic field of the more flux-rope-like CME. Further testing will show whether this is a persistent limitation of OSPREI, or specific to these two cases. The more flux-rope like CMEs tend to be the more extreme cases, with more potential of significant space weather effects, so successfully reproducing these is essential. The less structure CMEs tends to be weaker and of less concern, but also occur more frequently.

Since we have taken a forecasting approach, we have not shown  the absolute limitations of OSPREI. We have avoided fine-tuning of parameters as much as possible. While we expect that our ensemble seed inputs are close the the optimal values, it is entirely possible that we can improve our fit to observations, both in the corona and in situ, by slight adjustments in the inputs. Optimizing these results requires a careful exploration of parameter space, but understanding the best results OSPREI can produce in a scientific research setting is important as they will be a fundamental limitation on any prediction capabilities.

Beyond more thoroughly understanding the current version of OSPREI, we always seek to improve it. Fig. \ref{corr} shows a strong correlation between many outputs and the initial magnetic field strength of the background solar wind for both example CMEs. We also saw that while FIDO-SIT sheath model can reasonable reproduce the observed total $B$ magnitude and Kp, the prediction of the actual components leaves much to be desired as we can only currently assume a Parker Spiral orientation for the upstream direction. The background solar wind is one of the most under-developed components of the OSPREI suite and is an obvious area for improvement. Ideally, we want to have a more realistic background solar wind, but something that can be produced on a forecasting timescale. Recent improvements have allowed OSPREI to use a 1D profile for the background solar wind, which could easily be extracted from a steady-state solar wind forecast. Future work will test the effects of different backgrounds on the OSPREI results. 

Another factor that may be important is solar evolution on short time scales. We use a synchronic magnetogram that has been updated to use the most recent information around the central meridian, but this will not capture any evolution rotating around near the eastern limb. For both CMEs, we see active regions in the east. Our magnetograms will not have the most up-to-date information on them, only whatever was observed when those longitudes were last near the central meridian. Changes in the active regions may have important effects on the global PFSS structure. These effects need to be quantified in future work, as well as developing a strategy on how best to approach this issue in an operational setting. 

We also need to develop and test systematic methods for determining our ``difficult'' inputs, the CME mass, perpendicular angular width, and initial magnetic field strength and temperature. Methods exists for reconstructing, or at least estimating, all four values. Some of these are empirical relations that may have high ranges of uncertainty. Others are more robust, physics-driven approaches, such as deriving the mass from the white light brightness excess or estimating the reconnected flux based on the EUV signatures. Further study is needed to determine the effects of the different methods and develop approaches for both forecasting and producing scientific results.

As described here, OSPREI requires a coronal reconstruction of the CME to obtain the angular width and speed (and doublecheck that ForeCAT is giving reasonable results). This means it would likely take several hours to get the model started after the CME leaves the corona, but this should still give several days warning for most CMEs. If we were to use robust relationships between flare or active region properties and the CME angular width and size then we could potentially make the prediction without the coronal reconstruction. Many empirical relations exists, but have not yet been tested in the OSPREI suite. Interestingly, in the case of using active region properties, we could even run hypothetical cases before an eruption occurs. These would not likely be the most precise predictions, but could give very early warning of the general expectation, including whether or not much southward magnetic field may occur.

\section{Conclusion}
We have presented the new CME forecasting suite OSPREI, which represents the combination of previously established models and includes standardized, user-friendly visualizations. To demonstrate OSPREI's capabilities we have applied it to the CMEs observed in the corona on 2021 May 09 and 2021 April 22, detailing how we determine the inputs for each case. We mimic a forecasting approach, using the best information that would be available in a real-time scenario, rather than fine-tuning inputs to optimize the output for a detailed scientific study. 

OSPREI first performs a ForeCAT simulation of the coronal deflection and rotation of each CME. This gives any changes in the CME latitude, longitude, and orientation during the coronal propagation. This information is essential for determining whether or not a CME will impact a given location.

This positional information is then propagated to an arrival time model, ANTEATR. ANTEATR includes the basic drag between the CME and background solar wind, analogous to most current arrival time models, but also includes magnetic and thermal forces. In addition to simulating the CME's propagation, giving the arrival time and the CME's bulk and expansion velocities, ANTEATR determines the expansion and deformation of a CME, giving the angular widths and aspect ratios. Since ANTEATR depends on the internal properties the model also yields the expected temperature, magnetic field, and density of the CME at the time of impact.

The final component, FIDO, takes the positional information from ForeCAT and CME arrival time and properties from ANTEATR and determines an in situ magnetic field and velocity profile as it passes over a synthetic spacecraft. FIDO simulates not only the profile related to the flux rope or magnetic obstacle portion of a CME, but can also provide an estimate of the behavior at the CME-driven shock and within the sheath. FIDO uses the magnetic field and velocity profiles to estimate the Kp index over time.

The individual models had been designed for efficiency. Now that they are fully coupled, full Sun-to-Earth CME simulations can be run in about half a minute per case. This allows us to run ensembles so that we can better understand the range of possibilities in the predictions. Applying this ensemble methodology to two different CMEs highlights some of the varying behavior we may see. For example, the ForeCAT results show little deflection or rotation for the 2021 May 09 CME and the final latitude and longitude of all ensemble members are within 2$^{\circ}$. In contrast, the 2021 April 22 CME exhibits noticeable deflection and the magnitude varies between ensemble members, causing the final latitudes and longitudes to vary by 10$^{\circ}$.

In addition to providing a new modeling framework, we present a set of visualizations that can be automatically generated for OSPREI results and have been optimized to facilitate forecasts. Some of these visualizations represent novel approaches for the realm of forecasting. For example, we can visualize the percentage chance of each in situ property ($B$, $v$, $Kp$) having specific values as a function of time. We produce spatial maps of the average CME properties at the latitudes and longitudes around the Earth or satellite of interest, providing information on what would change if there was a slight change in the location of impact. Once OSPREI begins the research-to-operation transition this information could greatly improve our ability to forecast CME-driven space weather, both at Earth and beyond.

OSPREI also provides exciting opportunities for scientific research where one may want to dive deep into the full ensemble details. We can immediately visualize the correlation between inputs and outputs, confirming expected trends and unveiling unexpected ones. Additionally, we can closely examine the outlier cases and better understand what input or combination of inputs causes the abnormal behavior. In some cases, it may just result from inappropriate inputs that result in borderline unstable forces during interplanetary propagation, but this still provides physics-based information on the plausible range of hard to measure values such as a CME's coronal magnetic field strength, which can improve our scientific understanding, in general. 

\acknowledgments
CK is supported by the National Aeronautics and Space Administration under Grant 80NSSC19K0274 issued through the Heliophysics Guest Investigators Program and by the National Aeronautics and Space Administration under Grant 80NSSC19K0909 issued through the Heliophysics Early Career Investigators Program. 

The full OSPREI code is archived through Zenodo at \url{https://zenodo.org/badge/latestdoi/236822923} with DOI 10.5281/zenodo.5507441 and is available at GitHub at \url{https://github.com/ckay314/OSPREI}. SDO/HMI magnetograms are available at \url{http://jsoc.stanford.edu/ajax/lookdata.html?ds=hmi.Mrdailysynframe_720s}. The DONKI entry for the 2021 May 09 CME is available at \url{https://kauai.ccmc.gsfc.nasa.gov/DONKI/view/CMEAnalysis/16870/1} and the 2021 April 22 CME at available at \url{https://kauai.ccmc.gsfc.nasa.gov/DONKI/view/CME/16772/1} with the corresponding shock arrival at \url{https://kauai.ccmc.gsfc.nasa.gov/DONKI/view/IPS/16784/1}. SDO composite imagery is available from \url{https://sdo.gsfc.nasa.gov/data/aiahmi/}. The STEREO coronagraph data is available at \url{https://secchi.nrl.navy.mil/secchi_flight/images}. OMNI data are available at \url{https://omniweb.gsfc.nasa.gov/}.



%
%

%
%
%
%
%

\end{document}